\documentclass[aps,twocolumn,prd,preprintnumbers,superscriptaddress,nobibnotes,floatfix,longbibliography,nofootinbib]{revtex4-2}

\pdfoutput=1
\usepackage{amsmath,amsfonts,amssymb,mathrsfs,graphicx,color,longtable}
\usepackage{hyperref}
\usepackage{bm}
\usepackage{array}
\usepackage{color}
\usepackage{enumitem}
\usepackage[explicit]{titlesec}
\usepackage{ulem}
\usepackage[utf8]{inputenc}
\usepackage{float}  
\usepackage{multirow,bigdelim}
\usepackage{blindtext}
\usepackage{rotating}
\usepackage[ddmmyy,24hr]{datetime}

\usepackage[TABBOTCAP,tight]{subfigure}
\newcolumntype{C}[1]{>{\centering\let\newline\\\arraybackslash\hspace{4pt}}m{#1}}

\graphicspath{{figures/}}

\begin{document}

\title{Gallium Anomaly: Critical View from the Global Picture of $\nu_{e}$ and $\bar\nu_{e}$ Disappearance }

\author{C. Giunti}
\email{carlo.giunti@to.infn.it}
\affiliation{Istituto Nazionale di Fisica Nucleare (INFN), Sezione di Torino, Via P. Giuria 1, I--10125 Torino, Italy}

\author{Y.F. Li}
\email{liyufeng@ihep.ac.cn}
\affiliation{Institute of High Energy Physics, Chinese Academy of Sciences, Beijing 100049, China}
\affiliation{School of Physical Sciences, University of Chinese Academy of Sciences, Beijing 100049, China}

\author{C.A. Ternes}
\email{ternes@to.infn.it}
\affiliation{Istituto Nazionale di Fisica Nucleare (INFN), Sezione di Torino, Via P. Giuria 1, I--10125 Torino, Italy}
\affiliation{Dipartimento di Fisica, Universit\`a di Torino, via P. Giuria 1, I–10125 Torino, Italy}

\author{O.~Tyagi}
\email{oddhar89\_sps@jnu.ac.in}
\affiliation{Istituto Nazionale di Fisica Nucleare (INFN), Sezione di Torino, Via P. Giuria 1, I--10125 Torino, Italy}
\affiliation{School of Physical Sciences,  Jawaharlal Nehru University, New Delhi 110067, India}

\author{Z. Xin}
\email{xinzhao@ihep.ac.cn}
\affiliation{Institute of High Energy Physics,
Chinese Academy of Sciences, Beijing 100049, China}
\affiliation{School of Physical Sciences, University of Chinese Academy of Sciences, Beijing 100049, China}

\date{\dayofweekname{\day}{\month}{\year} \ddmmyydate\today, \currenttime}

\begin{abstract}
The significance of the Gallium Anomaly, from the BEST, GALLEX, and SAGE radioactive source experiments, is quantified using different theoretical calculations of the neutrino detection cross section, and its explanation due to neutrino oscillations is compared with the bounds from the analyses of reactor rate and spectral ratio data, $\beta$-decay data, and solar neutrino data.  
In the 3+1 active-sterile neutrino mixing scheme, the Gallium Anomaly is in strong tension with the individual and combined bounds of these data sets. In the combined scenario with all available data, the parameter goodness of fit is below 0.042\%, corresponding to a severe tension of 4--5$\sigma$, or stronger. Therefore, we conclude that one should pursue other possible solutions beyond short-baseline oscillations for the Gallium Anomaly. We also present a new global fit of $\nu_e$ and $\bar\nu_e$ disappearance data, showing that there is a 2.6--3.3$\sigma$ preference in favor of short-baseline oscillations,
which is driven by an updated analysis of reactor spectral ratio data.
\end{abstract}

\maketitle
\tableofcontents

%%%%%%%%%%%%%%%%%%%%%%%%%%%%%%%%%%%%%%%%%%%%%%%%%%%%%%%%

\section{Introduction}
\label{sec:intro}

The possible existence of light sterile neutrinos
is a hot topic of current research in high-energy physics.
It was motivated by anomalies found in
short-baseline neutrino oscillation experiments:
the Gallium Anomaly,
the Reactor Antineutrino Anomaly,
and the LSND and MiniBooNE anomalies
(see the reviews in Refs.~\cite{Gariazzo:2015rra,Gonzalez-Garcia:2015qrr,Giunti:2019aiy,Diaz:2019fwt,Boser:2019rta,Dasgupta:2021ies}).
Most puzzling is the large Gallium Anomaly,
which has been recently revived by the results of the
BEST experiment~\cite{Barinov:2021asz,Barinov:2022wfh}.
In this paper we discuss the status of
short-baseline $\nu_{e}$ and $\bar\nu_{e}$ disappearance
and we compare the neutrino oscillation explanation of the Gallium Anomaly
with the constraints from other experiments.

The standard paradigm in the phenomenology of massive neutrinos
is the three-neutrino mixing scheme in which
the three well known active flavor neutrinos
$\nu_{e}$,
$\nu_{\mu}$, and
$\nu_{\tau}$
take part in the weak interactions of the Standard Model
and are unitary superpositions of three massive neutrinos
$\nu_{1}$,
$\nu_{2}$, and
$\nu_{3}$
with respective masses
$m_{1}$,
$m_{2}$, and
$m_{3}$.
The two independent squared-mass differences
$\Delta m_{21}^2 \approx 7.4 \times 10^{-5} \, \text{eV}^2$
and
$|\Delta m_{31}^2| \approx 2.5 \times 10^{-3} \, \text{eV}^2$
(with $\Delta m_{kj}^2 \equiv m_{k}^2 - m_{j}^2 $)
generate the oscillations observed in
solar, atmospheric and long-baseline neutrino oscillation experiments
(see e.g., Ref.~\cite{Workman:2022ynf}
and the recent three-neutrino global analyses in
Refs.~\cite{deSalas:2020pgw,Esteban:2020cvm,Capozzi:2021fjo}).
Short-baseline (SBL) oscillations that could explain the
Reactor Antineutrino Anomaly
and the Gallium Anomaly
require the existence of at least one additional squared-mass difference
$\Delta m_{\text{SBL}}^2 \gtrsim 1 \, \text{eV}^2$.
In the minimal 3+1 scenario that we consider here
there is a non-standard massive neutrino $\nu_{4}$
with mass $m_{4} \gtrsim 1 \, \text{eV}$
which generates the short-baseline squared-mass difference
$\Delta m_{\text{SBL}}^2 \simeq \Delta m_{41}^2$.
In the flavor basis,
the new neutrino corresponds to a sterile neutrino
$\nu_{s}$
which does not
take part in the weak interactions of the Standard Model
(see the reviews in Refs.~\cite{Gariazzo:2015rra,Gonzalez-Garcia:2015qrr,Giunti:2019aiy,Diaz:2019fwt,Boser:2019rta,Dasgupta:2021ies}).
In this framework,
the effective short-baseline survival probability of electron neutrinos and antineutrinos relevant for
reactor and Gallium experiments
is given by
\begin{equation}
P_{ee}
\simeq
1 - \sin^2\!2\vartheta_{ee} \, \sin^2 \left(\frac{\Delta m_{41}^2 L}{4E}\right)
.
\label{Pee}
\end{equation}
The effective mixing angle $\vartheta_{ee}$
depends on the element $U_{e4}$ of the $4\times4$ mixing matrix $U$ through the relation
$ \sin^2\!2\vartheta_{ee} = 4 |U_{e4}|^2 ( 1 - |U_{e4}|^2 ) $.

Currently, the Reactor Antineutrino Anomaly is regarded to be resolved or, at least, diminished with the new refinements of reactor flux models~\cite{Berryman:2020agd,Giunti:2021kab}, but the Gallium Anomaly is reinforced by the new measurements of the BEST experiment~\cite{Barinov:2021asz,Barinov:2022wfh}. Therefore, it is desirable to pay special attention to the Gallium Anomaly, and look for possible viable solutions. In this work, we first evaluate the 
statistical significance of the Gallium Anomaly with different calculations of the neutrino detection cross section.
Then, we compare the neutrino oscillation explanation of the Gallium Anomaly with the bounds from several classes of $\nu_e$ and $\bar\nu_{e}$ disappearance experiments. In particular, we compare it with the above mentioned Reactor Antineutrino Anomaly and also with the results of an updated analysis of reactor spectral ratio data. Interestingly, the preference for short-baseline neutrino oscillations is reinforced when considering the newest spectral ratio data, as will be detailed below. We also consider data from tritium experiments and from experiments measuring solar neutrinos, before combining all data to a global $\nu_{e}$ and $\bar\nu_{e}$ disappearance fit. The consistency of a solution of the Gallium Anomaly with 3+1 active-sterile neutrino mixing will be compared and discussed at each step.

This work is organized as follows.
In Section~\ref{sec:gallium}, we discuss the analysis of Gallium data for several cross section models and we quantify the significance of the Gallium Anomaly. In Sections~\ref{sec:rates} and~\ref{sec:ratios} we compare the results of the Gallium experiments with those obtained from the analysis of reactor rate and spectral ratio data, respectively. Section~\ref{sec:rates-ratios} contains the combined analysis of all reactor data. In Section~\ref{sec:KATRIN} we detail the analysis procedure for the KATRIN data,
and in Section~\ref{sec:reactors-tritium}
we discuss the comparison of the regions of parameter space preferred by KATRIN and other $\beta$-decay experiments in combination with reactor data with those preferred by the Gallium data.
Section~\ref{sec:solar} discusses the updated solar neutrino bounds. Finally, in Section~\ref{sec:global} we combine all the data discussed in previous sections to a global 3+1 $\nu_e$ and $\bar\nu_{e}$ disappearance fit and we compare the results with those obtained from the Gallium data. We close with a discussion and a summary of our results in Section~\ref{sec:conclusions}.

%%%%%%%%%%%%%%%%%%%%%%%%%%%%%%%%%%%%%%%%%%%%%%%%%%%%%%%%

\section{The Gallium Anomaly}
\label{sec:gallium}

\begin{table*}
\centering
\begin{tabular}{lc|cc|cc|cc}
&
&
\multicolumn{2}{c|}{$^{51}\text{Cr}$}
&
\multicolumn{2}{c|}{$^{37}\text{Ar}$}
&
&
\\
\cline{3-4}
\cline{5-6}
Model
&
Method
&
$\sigma_{\text{tot}}$
&
$\delta_{\text{exc}}$
&
$\sigma_{\text{tot}}$
&
$\delta_{\text{exc}}$
&
$\overline{R}$
&
GA
\\
\hline
Ground State~\cite{Semenov:2020xea}
&
$ T_{1/2}({}^{71}\text{Ge}) $
&
$ 5.539 \pm 0.019 $
&
$-$
&
$ 6.625 \pm 0.023 $
&
$-$
&
$0.844 \pm 0.031$
&
$5.0\sigma$
\\
Bahcall (1997)~\cite{Bahcall:1997eg}
&
$ {}^{71}\text{Ga} (p,n) {}^{71}\text{Ge} $
&
$ 5.81 \pm 0.16 $
&
$4.7\%$
&
$ 7.00 \pm 0.21 $
&
$5.4\%$
&
$0.802 \pm 0.037$
&
$5.4\sigma$
\\
Haxton (1998)~\cite{Haxton:1998uc}
&
{Shell Model}
&
$ 6.39 \pm 0.65 $
&
$13.3\%$
&
$ 7.72 \pm 0.81 $
&
$14.2\%$
&
$0.703 \pm 0.078$
&
$3.8\sigma$
\\
Frekers et al. (2015)~\cite{Frekers:2015wga}
&
$ {}^{71}\text{Ga} ({}^{3}\text{He},{}^{3}\text{H}) {}^{71}\text{Ge} $
&
$ 5.92 \pm 0.11 $
&
$6.4\%$
&
$ 7.15 \pm 0.14 $
&
$7.3\%$
&
$0.788 \pm 0.032$
&
$6.5\sigma$
\\
Kostensalo et al. (2019)~\cite{Kostensalo:2019vmv}
&
{Shell Model}
&
$ 5.67 \pm 0.06 $
&
$2.3\%$
&
$ 6.80 \pm 0.08 $
&
$2.6\%$
&
$0.824 \pm 0.031$
&
$5.6\sigma$
\\
Semenov (2020)~\cite{Semenov:2020xea}
&
$ {}^{71}\text{Ga} ({}^{3}\text{He},{}^{3}\text{H}) {}^{71}\text{Ge} $
&
$ 5.938 \pm 0.116 $
&
$6.7\%$
&
$ 7.169 \pm 0.147 $
&
$7.6\%$
&
$0.786 \pm 0.033$
&
$6.6\sigma$
\\
\hline
\end{tabular}

\caption{\label{tab:gallium_cross_sections}
$\nu_{e} + {}^{71}\text{Ga} \to {}^{71}\text{Ge} + e^{-}$
cross sections in units of $10^{-45} \, \text{cm}^2$
and the corresponding relative contributions
$\delta_{\text{exc}}$
of the transitions to the excited states.
Also shown are the average ratio
$\overline{R}$
of observed and predicted events
and the statistical significance of the corresponding Gallium Anomaly.
}
\end{table*}

\begin{figure*}
\centering
\subfigure[]{ \label{fig:gallium-bahcall-a}
\includegraphics[width=0.48\linewidth]{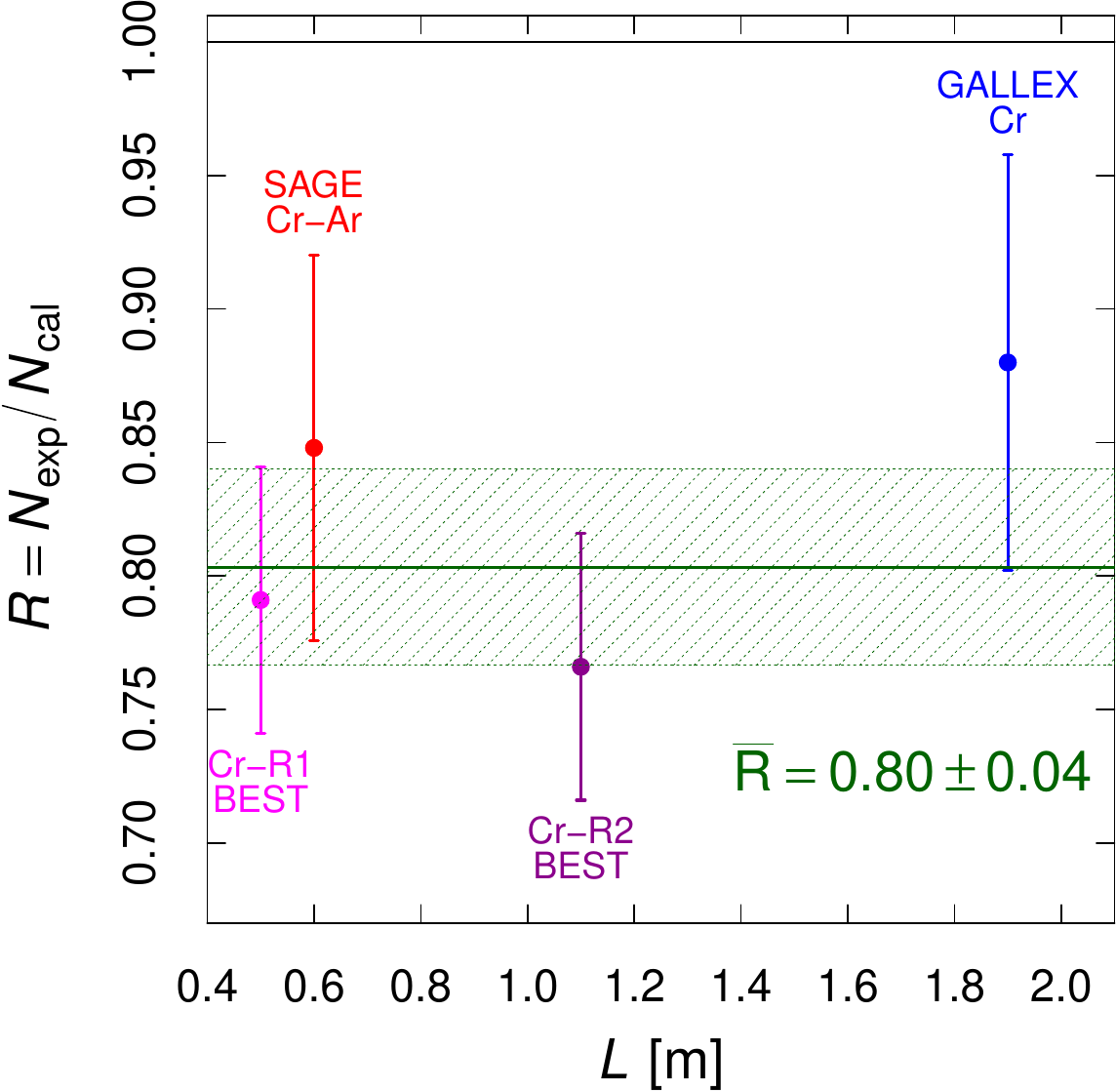}
}
\subfigure[]{ \label{fig:gallium-bahcall-b}
\includegraphics[width=0.48\textwidth]{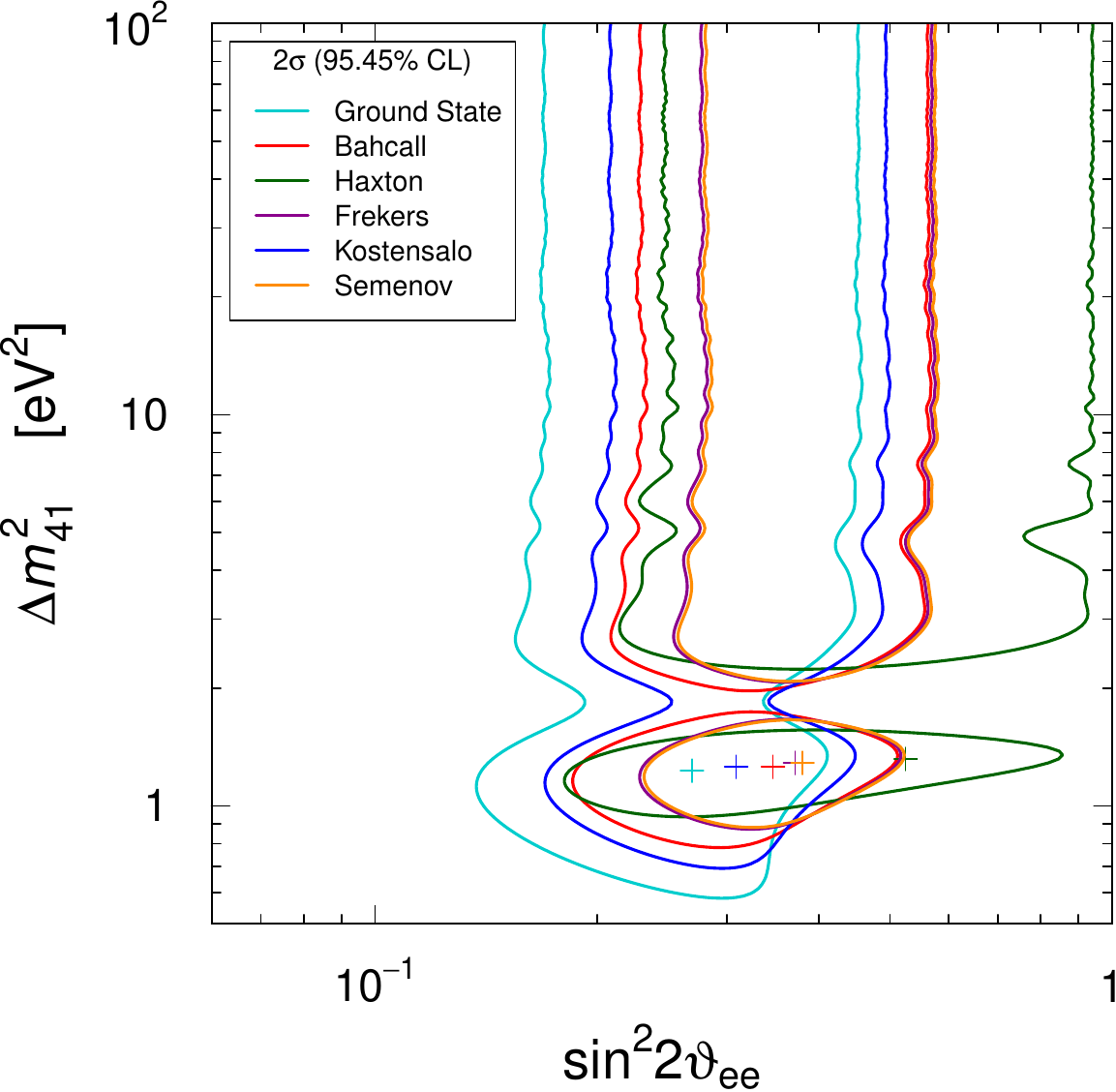}
}
\caption{\label{fig:gallium-bahcall}
\subref{fig:gallium-bahcall-a}
Ratios of observed and predicted event rates in the Gallium source experiments
obtained using the Bahcall cross sections~\cite{Bahcall:1997eg}.
\subref{fig:gallium-bahcall-b}
Contours enclosing the $2\sigma$ allowed regions in the
($\sin^2\!2\vartheta_{ee},\Delta{m}^2_{41}$)
plane obtained from the Gallium data with the different cross sections in
Tab.~\ref{tab:gallium_cross_sections}.
The best-fit points are indicated by crosses.
Note that the Frekers (magenta) and Semenov (orange) contours
are almost superimposed.
}
\end{figure*}

The Gallium Anomaly was originally~\cite{Abdurashitov:2005tb,Laveder:2007zz,Giunti:2006bj}
a deficit of events observed in the
GALLEX~\cite{GALLEX:1994rym,GALLEX:1997lja,Kaether:2010ag}
and
SAGE~\cite{Abdurashitov:1996dp,SAGE:1998fvr,Abdurashitov:2005tb,SAGE:2009eeu}
source experiments
aimed at testing the solar neutrino detection done in these experiments
through the process
$\nu_e + {}^{71}\text{Ga} \to e^- + {}^{71}\text{Ge}$.
Two source experiments have been done by the GALLEX collaboration
using an intense artificial $^{51}$Cr radioactive source
placed inside the detector.
This source
emitted electron neutrinos through the electron capture (EC) process
$e^- + {}^{51}\text{Cr} \to {}^{51}\text{V} + \nu_e$.
The SAGE collaboration performed a source experiment
with a $^{51}$Cr radioactive source
and another with a $^{37}$Ar radioactive source,
which emitted electron neutrinos through the EC process
$e^- + {}^{37}\text{Ar} \to {}^{37}\text{Cl} + \nu_e$.
The deficits of the observed rates with respect to the rates
calculated from the well-measured activity of the sources
and different cross sections for the detection process
(see Tab.~\ref{tab:gallium_cross_sections})
have been discussed in many papers
(see the reviews in Refs.~\cite{Gariazzo:2015rra,Gonzalez-Garcia:2015qrr,Giunti:2019aiy,Diaz:2019fwt,Boser:2019rta,Dasgupta:2021ies}).
The deficits obtained using the earliest Bahcall cross section~\cite{Bahcall:1997eg}
are illustrated in Fig.~\ref{fig:gallium-bahcall-a},
where we plotted the ratios of observed and predicted rates
versus the average path lengths of the neutrinos
(about $1.9~\text{m}$ for GALLEX and $0.6~\text{m}$ for SAGE).
The figure shows also the results of the recent
BEST source experiment~\cite{Barinov:2021asz,Barinov:2022wfh},
where a larger deficit was observed,
confirming the Gallium Anomaly.
The two BEST values in Fig.~\ref{fig:gallium-bahcall-a}
show the ratio of observed and predicted rates in the two nested
${}^{71}\text{Ga}$ volumes of the experiment,
which correspond to average neutrino path lengths of about
$0.5~\text{m}$ and $1.1~\text{m}$~\cite{Barinov:2021asz,Barinov:2022wfh}.
Although the ratios shown in Fig.~\ref{fig:gallium-bahcall-a}
exhibit some variation with distance,
we cannot see a clear oscillatory behavior.
Taking into account the large error bars
and a 2.8\% correlated systematic uncertainty of the Bahcall cross section,
the data can be fitted with a constant average ratio
$\overline{R} = 0.80 \pm 0.04$.
We considered the
systematic uncertainty of the $^{51}$Cr cross section
as correlated among all the experiments
and we added in quadrature the small residual systematic uncertainty
of the $^{37}$Ar cross section
to the uncertainty of the SAGE $^{37}$Ar measurement.

From the absence of a clear oscillatory pattern as a function of distance in 
Fig.~\ref{fig:gallium-bahcall-a}
and, in particular,
the quasi-equality of the two BEST measurements at different distances,
it follows that there is no smoking-gun evidence of
oscillations in the Gallium data.
After the BEST measurements,
the Gallium Anomaly is still an anomaly
based on the absolute comparison of observed and predicted rates,
as it was when only the GALLEX and SAGE data were available.
Therefore,
a crucial role is played by the theoretical detection cross section,
for which there are the different model calculations listed in
Tab.~\ref{tab:gallium_cross_sections}.
The difference between these cross section models is the
contribution to the cross section coming from the transitions
from the ground state of ${}^{71}\text{Ga}$
to excited states of ${}^{71}\text{Ge}$.
As shown in the first line of Tab.~\ref{tab:gallium_cross_sections},
the ground-state to ground-state cross section is known with a very small
uncertainty from the measured lifetime of ${}^{71}\text{Ge}$~\cite{Bahcall:1997eg}.
The table shows the relative contributions
$\delta_{\text{exc}}$
of the transitions to the excited states
in the different calculations of the cross section.
These relative contributions vary from about 2-3\%
in the Shell Model calculation of
Kostensalo et al.~\cite{Kostensalo:2019vmv}
to about 13-14\%
in the Shell Model calculation of
Haxton~\cite{Haxton:1998uc}.

Table~\ref{tab:gallium_cross_sections}
shows also the values of the average ratio
$\overline{R}$
of observed and predicted events
obtained with the different cross section models
and the statistical significance of the corresponding Gallium Anomaly.
One can see that the statistical significance of the Gallium Anomaly
is large, about $5$-$6\sigma$,
for all the cross section models,
except for the Haxton cross section model,
because of its large uncertainties, in spite of the larger Haxton cross
sections.
Let us however note that
the Shell Model cross section of Haxton~\cite{Haxton:1998uc},
calculated in 1998,
should be considered as superseded by the more recent
Shell Model cross section of Kostensalo et al.~\cite{Kostensalo:2019vmv},
which was calculated in 2019
using a state-of-the-art code and recently developed two-nucleon interactions.

In the first line of Tab.~\ref{tab:gallium_cross_sections}
(Ground State model)
we considered the possibility that the cross section is dominated by the
transition to the ground state of ${}^{71}\text{Ge}$,
with negligible contributions of the transitions to the excited states of ${}^{71}\text{Ge}$.
The value of the Ground State cross section is taken from the recent estimate in
Ref.~\cite{Semenov:2020xea}.
This is an extreme possibility that is justified by the
reliability of the cross section to the ground state of ${}^{71}\text{Ge}$
discussed above
and by the uncertainties of the cross sections
to the excited states of ${}^{71}\text{Ge}$,
which depend on the methods and assumptions of the different models.
The Ground State model corresponds to
the minimum possible value of the cross section
and gives the maximum possible value of the average ratio
$\overline{R}$
of observed and predicted events.
As one can see from Tab.~\ref{tab:gallium_cross_sections},
even in this extreme case the value of $\overline{R}$
is $5.0\sigma$ below one.
This is a strong signal of the seriousness of the Gallium Anomaly.

The Gallium Anomaly can be explained by short-baseline neutrino oscillations
in the 3+1 active-sterile neutrino mixing framework
described briefly in Section~\ref{sec:intro}.
Fig.~\ref{fig:gallium-bahcall-b}
shows the $2\sigma$ allowed regions in the
($\sin^2\!2\vartheta_{ee},\Delta{m}^2_{41}$)
plane obtained from the analyses
of the Gallium data with the different cross sections in
Tab.~\ref{tab:gallium_cross_sections}. 
One can see that in all cases there is a clear indication
of a relatively large value of $\sin^2\!2\vartheta_{ee}$
for $\Delta m_{41}^2 \gtrsim 0.6 \, \text{eV}^2$.
The minimum $2\sigma$ value of $\sin^2\!2\vartheta_{ee}$
is about $0.14$ in the case of the Ground State model.
This large value of $\sin^2\!2\vartheta_{ee}$ is required to produce
an oscillation amplitude that is sufficient
to explain the deficit represented by the
Ground State value of $\overline{R}$
in Tab.~\ref{tab:gallium_cross_sections}
at the $2\sigma$ level.
The smaller values of $\overline{R}$
for the other cross section models
correspond to allowed regions that cover larger values of
$\sin^2\!2\vartheta_{ee}$.
The Kostensalo Shell Model cross section requires values of $\sin^2\!2\vartheta_{ee}$
that are only slightly larger than those of the Ground State cross section,
$\sin^2\!2\vartheta_{ee} \gtrsim 0.17$,
because of the small contribution of the transitions
to the excited states of ${}^{71}\text{Ge}$ in this cross section model.
The older Shell Model cross section of Haxton,
which has much larger contributions of the transitions
to the excited states of ${}^{71}\text{Ge}$,
gives a broad allowed region that spans values of $\sin^2\!2\vartheta_{ee}$
from about $0.18$ to about $0.95$,
because of the large cross section uncertainties shown in
Tab.~\ref{tab:gallium_cross_sections}.
The Bahcall cross section implies
$\sin^2\!2\vartheta_{ee} \gtrsim 0.17$,
and the almost-equal Frekers and Semenov cross sections give
$\sin^2\!2\vartheta_{ee} \gtrsim 0.23$. Our results on the significance of the Gallium anomaly are in reasonable agreement with the results obtained in Ref.~\cite{Berryman:2021yan}.

In the following,
when we will compare the Gallium Anomaly with the results of other experiments,
we will consider only the following four cross section models:

\begin{description}

\item[Ground State]
This is a significant extreme case as discussed above.

\item[Bahcall]
It is the first and most used cross section model~\cite{Bahcall:1997eg}.

\item[Kostensalo]
It is the most recent Shell Model calculation~\cite{Kostensalo:2019vmv},
which supersedes the much older Haxton calculation~\cite{Haxton:1998uc},
as discussed above.

\item[Semenov]
It is a recent~\cite{Semenov:2020xea} reevaluation of the cross sections
which uses the
$ {}^{71}\text{Ga} ({}^{3}\text{He},{}^{3}\text{H}) {}^{71}\text{Ge} $
data of
Frekers et al.~\cite{Frekers:2015wga}
and supersedes their evaluation of the cross section.

\end{description}

%%%%%%%%%%%%%%%%%%%%%%%%%%%%%%%%%%%%%%%%%%%%%%%%%%%%%%%%

\section{The reactor rates constraints}
\label{sec:rates}

\begin{figure}
\centering
\includegraphics[width=\linewidth]{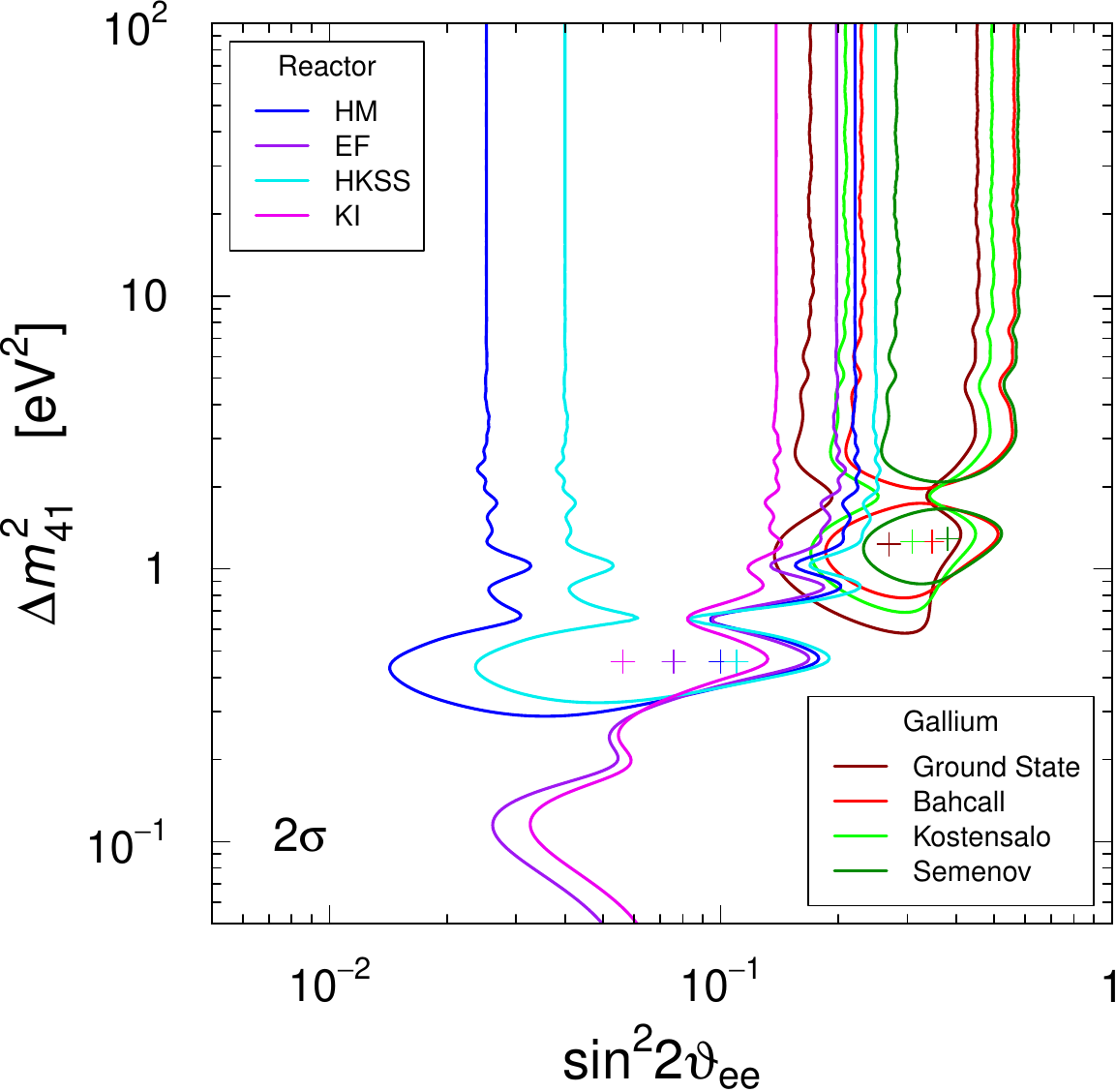}
\caption{\label{fig:see-gal-rat-2s}
Contours delimiting the $2\sigma$ allowed regions in the
($\sin^2\!2\vartheta_{ee},\Delta{m}^2_{41}$)
plane obtained from the Gallium data with different cross sections
compared with those obtained in Ref.~\cite{Giunti:2021kab}
from the analysis of short-baseline reactor rate data.
The best-fit points are indicated by crosses.
}
\end{figure}

The phenomenology of reactor neutrinos
focused on the Reactor Antineutrino Anomaly
in 2011~\cite{Mention:2011rk},
after the reevaluations of the predicted reactor antineutrino fluxes by
Mueller \textit{et al.}~\cite{Mueller:2011nm}
and
Huber~\cite{Huber:2011wv}
(HM model),
which led to a deficit of about $2.5\sigma$
of observed event rates in
short-baseline reactor neutrino experiments
with respect to the predictions.
Recent theoretical and experimental developments
led to new reactor flux models with enhanced
(HKSS model of
Hayen, Kostensalo, Severijns, and Suhonen~\cite{Hayen:2019eop})
or
diminished
(EF model of
Estienne, Fallot \textit{et al}~\cite{Estienne:2019ujo}
and
KI model of
Kopeikin \textit{et al.}~\cite{Kopeikin:2021ugh})
Reactor Antineutrino Anomaly.
In particular, the EF and KI flux models give
average ratios of observed and predicted events
that are only about $1\sigma$ smaller than unity,
hinting at the demise of the Reactor Antineutrino Anomaly~\cite{Berryman:2020agd,Giunti:2021kab}.
Here we use the results of the analysis in Ref.~\cite{Giunti:2021kab},
where it was shown that all the reactor flux models imply upper bounds
for the 3+1 active-sterile neutrino mixing parameter
$\sin^2\!2\vartheta_{ee}$
that are smaller than about $0.25$ at $2\sigma$.
This bound is in tension with the large values
of $\sin^2\!2\vartheta_{ee}$
which are required to explain the Gallium Anomaly,
as discussed in Section~\ref{sec:gallium}.

Figure~\ref{fig:see-gal-rat-2s} shows a
comparison of the $2\sigma$ allowed regions in the
($\sin^2\!2\vartheta_{ee},\Delta{m}^2_{41}$)
plane obtained from the analyses of reactor rates and Gallium data.
One can see that there is only a small overlap of the
reactor and Gallium regions for some models.
The largest overlaps occur for the
Ground State Gallium cross section model,
whose allowed region extends to the smallest values of $\sin^2\!2\vartheta_{ee}$
allowed by the Gallium Anomaly,
and for the HKSS reactor flux model,
which gives the largest Reactor Antineutrino Anomaly.

\begin{table*}
\centering
\begin{tabular}{|c|cc|cc|cc|cc|}
\cline{2-9}
\multicolumn{1}{c|}{}
&
\multicolumn{2}{c|}{HM}
&
\multicolumn{2}{c|}{HKSS}
&
\multicolumn{2}{c|}{EF}
&
\multicolumn{2}{c|}{KI}
\\
\multicolumn{1}{c|}{}
&
$\Delta\chi^{2}_{\text{PG}}$
&
$\text{GoF}_{\text{PG}}$
&
$\Delta\chi^{2}_{\text{PG}}$
&
$\text{GoF}_{\text{PG}}$
&
$\Delta\chi^{2}_{\text{PG}}$
&
$\text{GoF}_{\text{PG}}$
&
$\Delta\chi^{2}_{\text{PG}}$
&
$\text{GoF}_{\text{PG}}$
\\
\hline
Ground State
&
$7.2$
&
$2.8\%$
&
$5.4$
&
$6.8\%$
&
$9.1$
&
$1.1\%$
&
$11.9$
&
$0.26\%$
\\
\hline
Bahcall
&
$10.9$
&
$0.42\%$
&
$8.9$
&
$1.2\%$
&
$12.9$
&
$0.16\%$
&
$16.3$
&
$0.029\%$
\\
\hline
Kostensalo
&
$9.6$
&
$0.83\%$
&
$7.5$
&
$2.4\%$
&
$11.5$
&
$0.31\%$
&
$15.3$
&
$0.049\%$
\\
\hline
Semenov
&
$15.1$
&
$0.052\%$
&
$12.6$
&
$0.18\%$
&
$17.0$
&
$0.02\%$
&
$22.5$
&
$0.0013\%$
\\
\hline
\end{tabular}

\caption{ \label{tab:rat-gal-pgf}
$\chi^2$ difference
$\Delta\chi^{2}_{\text{PG}}$
of the parameter goodness of fit test~\cite{Maltoni:2003cu}
between the reactor rate data and the Gallium data
corresponding to four different reactor flux predictions
(HM, HKSS, EF, and KI)
discussed in the text
and four of the different Gallium detection cross sections
in Tab.~\ref{tab:gallium_cross_sections}.
The values of the corresponding parameter goodness of fit
$\text{GoF}_{\text{PG}}$
are calculated
with two degrees of freedom corresponding to the two
oscillation parameters
$\sin^2\!2\vartheta_{ee}$ and $\Delta{m}^2_{41}$
which are common in the reactor rates and Gallium data analyses.
}
\end{table*}

Table~\ref{tab:rat-gal-pgf}
shows the $\chi^2$ difference
$\Delta\chi^{2}_{\text{PG}}$
of the parameter goodness of fit test~\cite{Maltoni:2003cu}
between the reactor rates and the Gallium data
corresponding to the models in Fig.~\ref{fig:see-gal-rat-2s}.
Table~\ref{tab:rat-gal-pgf} shows also the corresponding
values of the parameter goodness of fit
$\text{GoF}_{\text{PG}}$,
which quantifies the tension between the fits of
reactor rates and Gallium data in the different models.
Considering the extreme case of Ground State Gallium cross section model,
the largest tension is obtained with the
KI reactor flux model ($\text{GoF}_{\text{PG}}=0.26\%$).
Indeed, one can see from Fig.~\ref{fig:see-gal-rat-2s}
that the corresponding $2\sigma$ allowed regions
have only a very marginal overlap for
$\Delta m_{41}^2 \simeq 1 \, \text{eV}^2$.

The more realistic
Bahcall, Kostensalo, and Semenov
cross section models give larger tensions between
the reactor rates and Gallium data.
If we consider as severe a tension with a $\text{GoF}_{\text{PG}}$
smaller than $1\%$,
for the Bahcall and Kostensalo cross section models there is a severe tension
between Gallium data and reactor rates for all reactor flux models,
except the HKSS model, while for the Semenov cross section model there is a severe tension for all of the reactor flux models.
One can also notice that the KI reactor flux model gives
the maximal tension for all Gallium cross section models
and it is always severe.
The EF reactor flux model gives a severe tension for all
the Gallium cross section models, except for the Ground State model,
where it is a marginal $1.1\%$.

Since the EF and KI reactor flux models are those that may have solved the
Reactor Antineutrino Anomaly~\cite{Berryman:2020agd,Berryman:2021yan,Giunti:2021kab}
and are currently believed to represent reliable replacements of the
standard HM model,
we conclude that the tension between
Gallium data and reactor rates
is a serious issue in the framework of 3+1 active-sterile neutrino oscillations.

Let us also emphasize that extending the 3+1 model with the introduction
of more sterile neutrinos does not help,
because any disappearance of neutrinos in Gallium experiments
must correspond to an antineutrino disappearance of the same size
in reactor experiments.
A possible loophole is a violation of the
CPT symmetry~\cite{Giunti:2010zs},
which implies the equality of the disappearance probabilities of neutrinos and
antineutrinos
(see, e.g., Ref.~\cite{Giunti-Kim-2007}).
Although this loophole may resolve the tension
between the
Gallium neutrino data and reactor antineutrino rates,
it cannot explain the tension between
the Gallium neutrino data and the solar neutrino bound
discussed in Section~\ref{sec:solar}.

%%%%%%%%%%%%%%%%%%%%%%%%%%%%%%%%%%%%%%%%%%%%%%%%%%%%%%%%

\section{Short-baseline reactor spectral ratios}
\label{sec:ratios}

Short-baseline oscillations
of reactor antineutrinos can be probed in a model-independent way
by comparing the rates or spectra measured at different distances
from the reactor antineutrino source.
This is the approach adopted by the recent
DANSS~\cite{DANSS:2018fnn,DANSS-ICHEP2022},
PROSPECT~\cite{PROSPECT:2018dtt,PROSPECT:2020sxr}, and
STEREO~\cite{STEREO:2018rfh,STEREO:2019ztb}
experiments.
Important results have been obtained also by the
NEOS collaboration~\cite{NEOS:2016wee}
and by a joint analysis of the RENO and NEOS collaborations~\cite{RENO:2020hva}.
In 2017~\cite{NEOS:2016wee} the
NEOS collaboration published the results of the comparison of NEOS data at about 24 m from the reactor source
with the prediction obtained from the neutrino flux measured in the Daya Bay experiment~\cite{DayaBay:2016ssb}
at a distance of about 550 m from the reactor source,
where the short-baseline oscillations are averaged.
Recently,
the RENO and NEOS collaborations
published a joint paper~\cite{RENO:2020hva}
with the results of the comparison of NEOS data
with the prediction obtained from the neutrino flux measured in the RENO experiment
at a distance of about 419 m from the reactor source,
which is in the same reactor complex of NEOS.
This comparison allowed the RENO and NEOS collaborations to reduce the systematic uncertainties
with respect to the NEOS/Daya Bay analysis.

In the following we consider both the
NEOS/Daya Bay
and
NEOS/RENO data,
because it is obviously unknown which one of the
Daya Bay and RENO neutrino spectra is more accurate.
For NEOS/Daya Bay we use the $\chi^2$ table
kindly provided by the NEOS collaboration,
as already done in Refs.~\cite{Gariazzo:2017fdh,Gariazzo:2018mwd,Giunti:2019fcj,Giunti:2020uhv}.
For NEOS/RENO we performed a fit using the information in the data release of Ref.~\cite{RENO:2020hva}.
We verified that our fit reproduces with good approximation the results published in Ref.~\cite{RENO:2020hva}.
For the DANSS experiment we use the $\chi^2$ table
kindly provided by the DANSS collaboration,
which corresponds to the latest results presented at the recent ICHEP 2022 conference~\cite{DANSS-ICHEP2022}.
For the PROSPECT experiment we performed a fit using the information in the data release of Ref.~\cite{PROSPECT:2020sxr}.
For the STEREO experiment we use the $\chi^2$ table in the data release of Ref.~\cite{STEREO:2019ztb}.

In this analysis we do not consider the controversial results
of the Neutrino-4 experiment:
the Neutrino-4 collaboration claimed a $2.9\sigma$
evidence of short-baseline neutrino oscillations
with large mixing ($\sin^2\!2\vartheta_{ee} = 0.36 \pm 0.12$)
at
$\Delta{m}^2_{41} = 7.3 \pm 1.17$~\cite{Serebrov:2020kmd}.
However, these results were criticized in Refs.~\cite{Danilov:2018dme,PROSPECT:2020raz,Danilov:2020rax,Giunti:2021iti}.
In particular,
the energy resolution of the detector was not taken into account
in the analysis of the Neutrino-4 collaboration,
as it is clear from a sentence in Ref.~\cite{Serebrov:2020kmd},
where they confuse energy binning
with taking into account the energy resolution of the detector. A combined analysis of spectral ratio data including Neutrino-4 (but not the latest DANSS data) has been performed in Ref.~\cite{Berryman:2021yan}. 

\begin{figure*}
\centering
\subfigure[]{ \label{fig:neos17}
\includegraphics[width=0.48\linewidth]{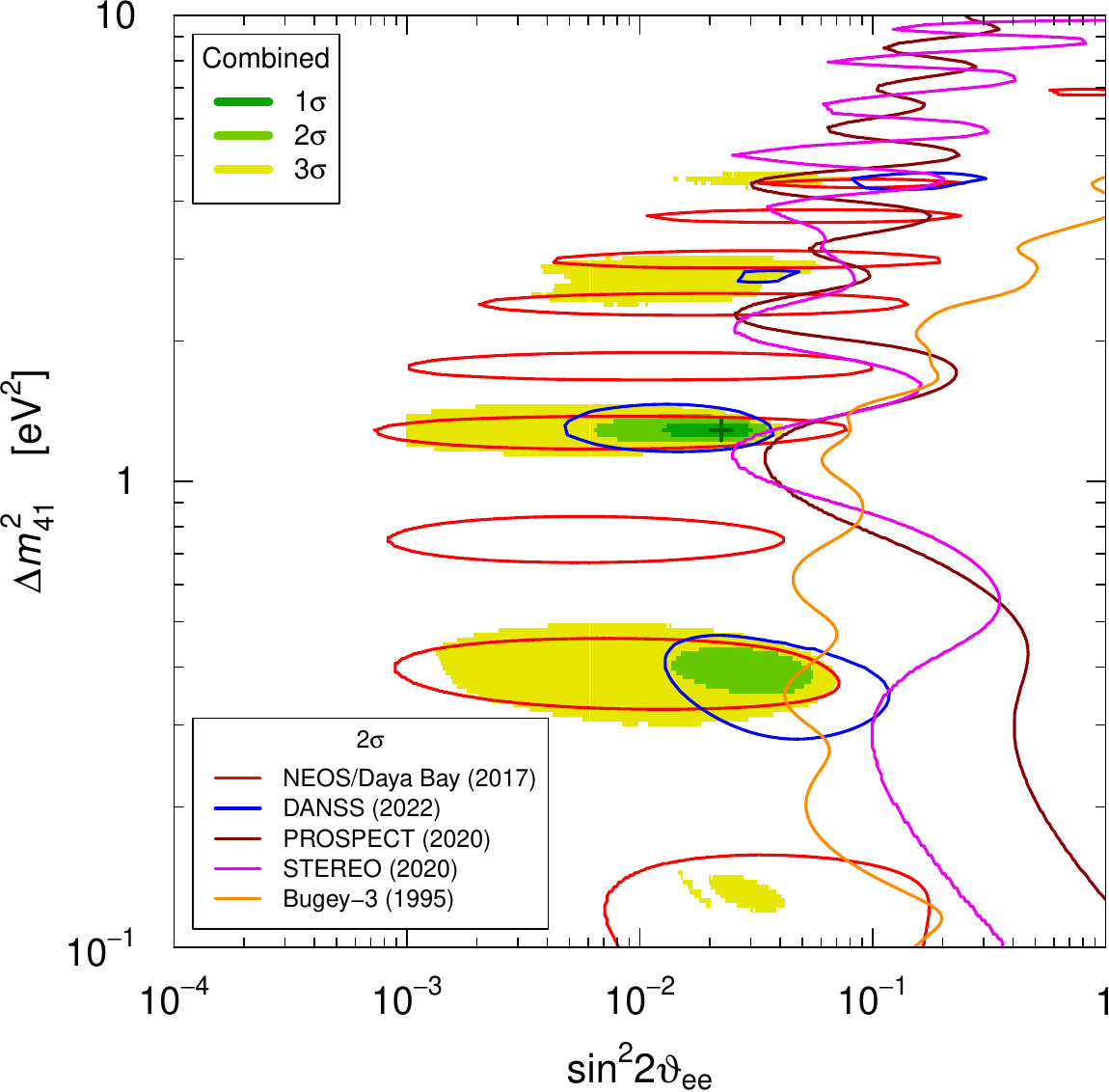}
}
\subfigure[]{ \label{fig:neos21}
\includegraphics[width=0.48\textwidth]{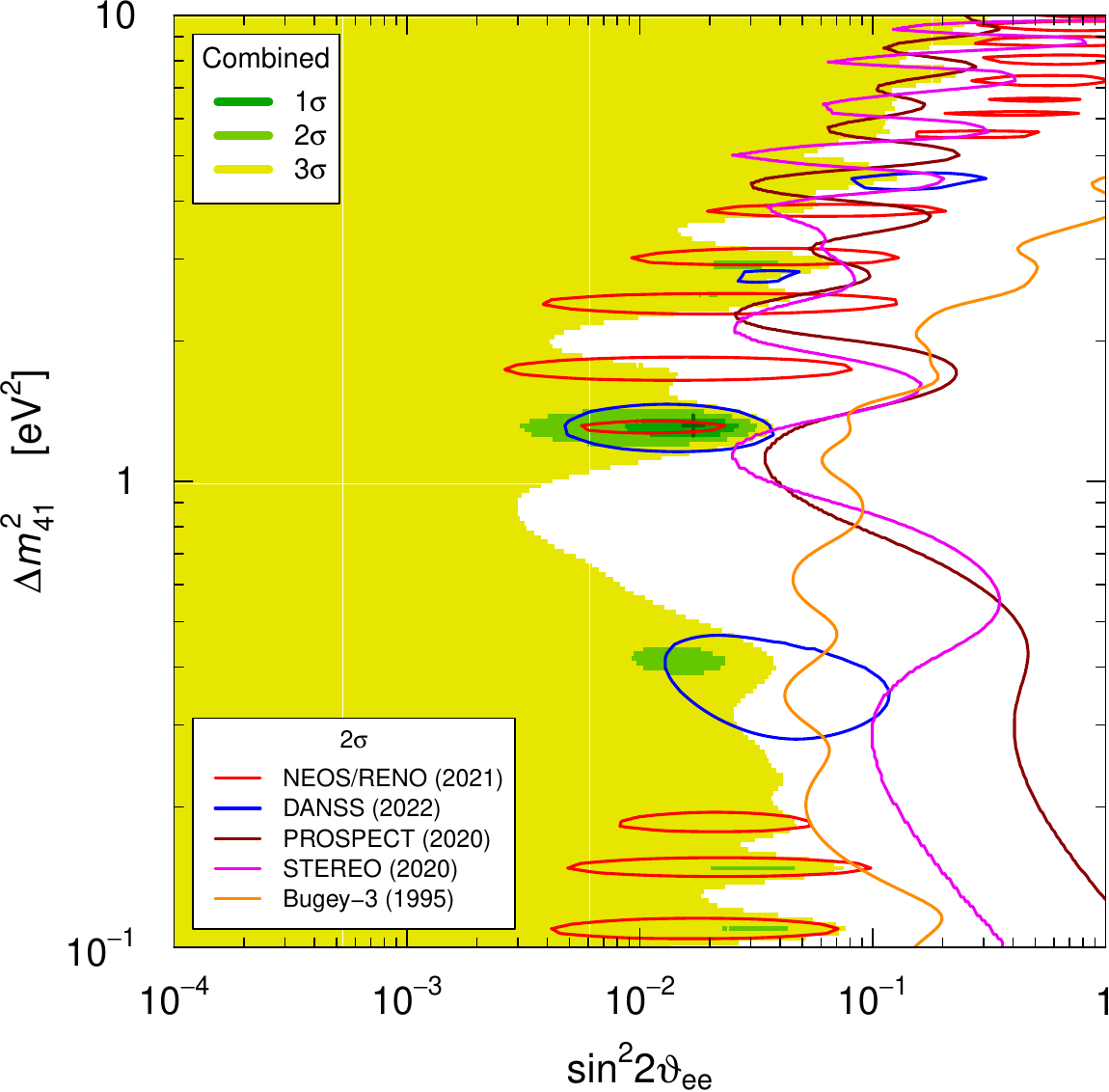}
}
\caption{\label{fig:spectral-ratios}
Allowed regions in the
($\sin^2\!2\vartheta_{ee},\Delta{m}^2_{41}$)
plane obtained from the combined analysis of the
\subref{fig:neos17}
NEOS/Daya Bay~\cite{NEOS:2016wee}
or
\subref{fig:neos21}
NEOS/RENO~\cite{RENO:2020hva}
spectral ratio data
with those of the
DANSS~\cite{DANSS-ICHEP2022},
PROSPECT~\cite{PROSPECT:2020sxr},
STEREO~\cite{STEREO:2019ztb}, and
Bugey-3~\cite{Declais:1995su}
experiments.
Also shown are the contours of the $2\sigma$ allowed regions of each experiments.
Closed contours surround the allowed region
and open contours exclude the region on the right.
}
\end{figure*}

Figure~\ref{fig:spectral-ratios}
shows the allowed regions in the
($\sin^2\!2\vartheta_{ee},\Delta{m}^2_{41}$)
plane obtained from the combined analysis of the
NEOS/Daya Bay
or
NEOS/RENO
spectral ratio data
with those of the other experiments.
For completeness,
besides the data of the recent
DANSS~\cite{DANSS-ICHEP2022},
PROSPECT~\cite{PROSPECT:2020sxr}, and
STEREO~\cite{STEREO:2019ztb}
experiments mentioned above,
we considered also the 40/50 m spectral ratio data of the old
Bugey-3 experiment~\cite{Declais:1995su},
as already done in Refs.~\cite{Gariazzo:2017fdh,Gariazzo:2018mwd,Giunti:2019fcj,Giunti:2020uhv}.
Figure~\ref{fig:spectral-ratios}
shows also the contours of the $2\sigma$ allowed regions of each experiments.
One can see that at $2\sigma$
PROSPECT,
STEREO, and
Bugey-3 produce only exclusion curves,
whereas
NEOS/Daya Bay,
NEOS/RENO, and
DANSS
yield closed contours.

The result of the combined fit is different when we consider
NEOS/Daya Bay
or
NEOS/RENO.
The combined fit with NEOS/Daya Bay (Fig.~\ref{fig:neos17})
results in a $3.1\sigma$ indication in favor of
short-baseline oscillations with best-fit parameters values
$\sin^2\!2\vartheta_{ee}=0.022$
and
$\Delta m_{41}^2 = 1.29 \, \text{eV}^2$.
One can see that the surrounding region is the locus of a remarkable overlap of relatively large
$2\sigma$ allowed regions of
NEOS/Daya Bay
and
DANSS,
which are not excluded by
the $2\sigma$ exclusion curves of
PROSPECT,
STEREO, and
Bugey-3.
In the combined fit with NEOS/RENO (Fig.~\ref{fig:neos21}),
the best-fit is approximately the same,
$\sin^2\!2\vartheta_{ee}=0.017$
and
$\Delta m_{41}^2 = 1.32 \, \text{eV}^2$,
but the indication in favor of
short-baseline oscillations is only of $2.6\sigma$.
Therefore, in Fig.~\ref{fig:neos21} the combined $3\sigma$ allowed region is not bounded on the left,
allowing a vanishing $\sin^2\!2\vartheta_{ee}$
which corresponds to the absence of short-baseline oscillations.

\begin{figure}
\centering
\includegraphics[width=\linewidth]{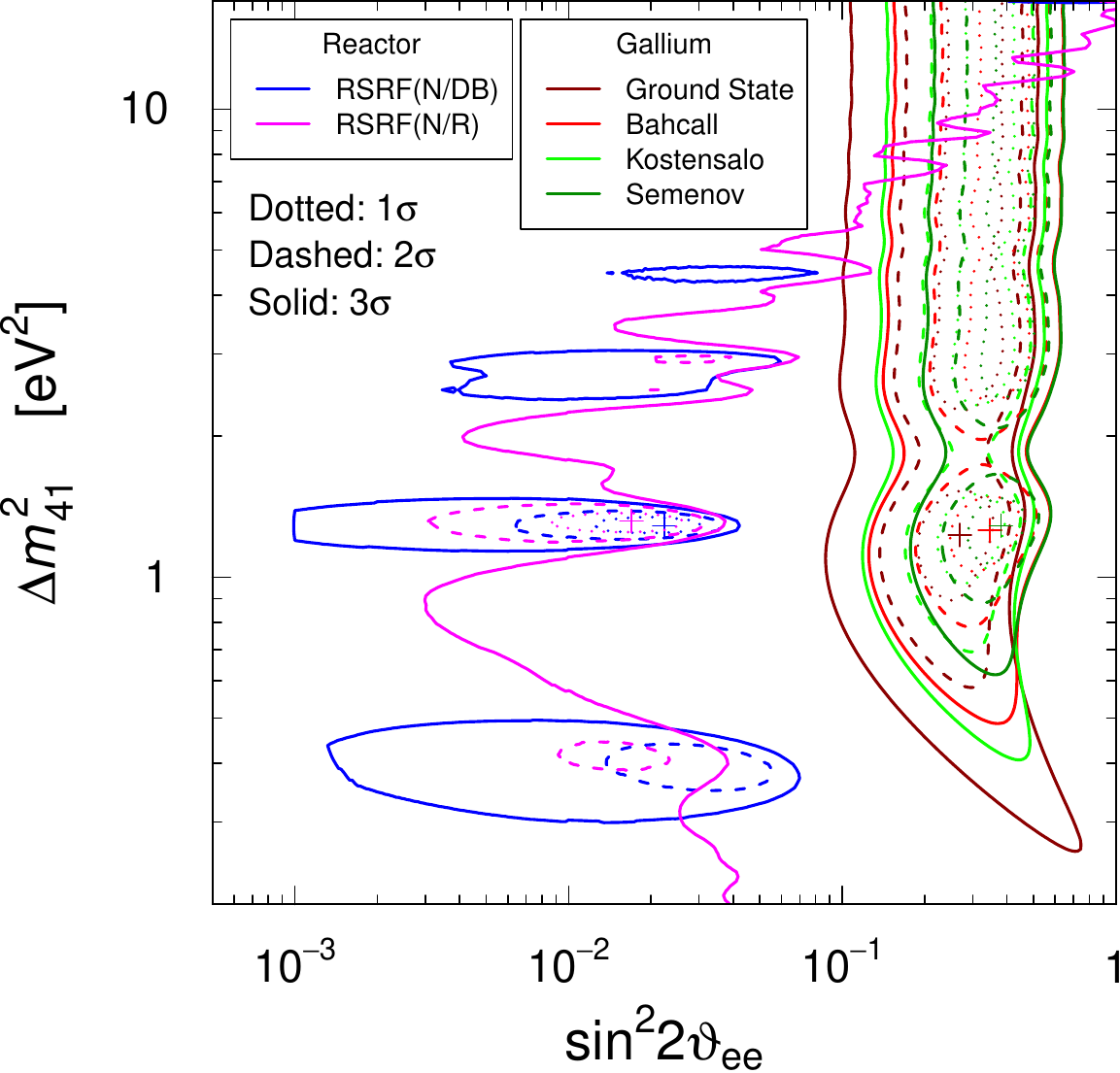}
\caption{\label{fig:see-gal-spe-sig}
Contours delimiting the
$1\sigma$,
$2\sigma$, and
$3\sigma$
allowed regions in the
($\sin^2\!2\vartheta_{ee},\Delta{m}^2_{41}$)
plane obtained from the Gallium data with different cross sections
compared with those obtained from
the two analyses of reactor spectral ratio data
discussed in the text.
The best-fit points are indicated by crosses.
}
\end{figure}

\begin{table*}
\centering
\begin{tabular}{|c|cc|cc|}
\cline{2-5}
\multicolumn{1}{c|}{}
&
\multicolumn{2}{c|}{RSRF(N/DB)}
&
\multicolumn{2}{c|}{RSRF(N/R)}
\\
\multicolumn{1}{c|}{}
&
$\Delta\chi^{2}_{\text{PG}}$
&
$\text{GoF}_{\text{PG}}$
&
$\Delta\chi^{2}_{\text{PG}}$
&
$\text{GoF}_{\text{PG}}$
\\
\hline
Ground State
&
$12.95$
&
$0.15\%$
&
$8.91$
&
$1.2\%$
\\
\hline
Bahcall
&
$12.86$
&
$0.16\%$
&
$8.74$
&
$1.3\%$
\\
\hline
Kostensalo
&
$12.91$
&
$0.16\%$
&
$8.89$
&
$1.2\%$
\\
\hline
Semenov
&
$12.88$
&
$0.16\%$
&
$8.70$
&
$1.3\%$
\\
\hline
\end{tabular}

\caption{ \label{tab:spe-gal-pgf}
$\chi^2$ difference
$\Delta\chi^{2}_{\text{PG}}$
of the parameter goodness of fit test~\cite{Maltoni:2003cu}
applied to the comparison of the neutrino oscillation fits of the reactor spectral ratio data and the Gallium data
with the four Gallium detection cross sections
in Tab.~\ref{tab:gallium_cross_sections}.
The values of the corresponding parameter goodness of fit
$\text{GoF}_{\text{PG}}$
are calculated
with two degrees of freedom corresponding to the two
common oscillation parameters
$\sin^2\!2\vartheta_{ee}$ and $\Delta{m}^2_{41}$.
The column titles RSRF(N/DB) and RSRF(N/R)
indicate, respectively,
the combined fits of NEOS/Daya Bay and NEOS/RENO data with those
of the other reactor spectral ratio experiments discussed in the text
(DANSS, PROSPECT, STEREO, and Bugey-3).
}
\end{table*}

Note that a more reliable estimate of the statistical significance
of the indications in favor of neutrino oscillations
would require Monte Carlo simulations,
because of the violation of Wilks'
theorem~\cite{Agostini:2019jup,Giunti:2020uhv,Coloma:2020ajw,Berryman:2021yan}.
However, these simulations are very difficult,
because they require detailed knowledge of all the experimental features
and huge computing times.
Therefore,
in this work we discuss only the results obtained assuming
the standard $\chi^2$ distribution,
as done in the great majority of publications.

In the following we discuss the compatibility of the results of the fits
of reactor spectral ratio data
with the neutrino oscillation interpretation of the Gallium data
and we perform combined fits with other experimental results.
For convenience, we introduce the following notation:
\begin{description}
\item[RSRF(N/DB)]
combined reactor spectral ratio fit of the
NEOS/Daya Bay, DANSS, PROSPECT, STEREO, and Bugey-3
data.
\item[RSRF(N/R)]
combined reactor spectral ratio fit of the
NEOS/RENO, DANSS, PROSPECT, STEREO, and Bugey-3
data.
\end{description}

As shown in Fig.~\ref{fig:see-gal-spe-sig},
these combined fits of the data of reactor spectral ratio experiments
favor short-baseline neutrino oscillations
at values of $\Delta{m}^2_{41}$ which are compatible with the
neutrino oscillation interpretation of the Gallium data
discussed in Section~\ref{sec:gallium},
but the required values of the mixing angle are smaller.
Therefore, there is a tension between the results of
reactor spectral ratio experiments
and those of the Gallium experiments.
This tension is quantified by the values of the parameter goodness of fit
in Tab.~\ref{tab:spe-gal-pgf}.
For all the four Gallium detection cross section models,
the parameter goodness of fit is well below 1\% for RSRF(N/DB),
whereas it is slightly above 1\% for RSRF(N/R).
The larger compatibility of RSRF(N/R)
with the Gallium data may seem contradictory with the smaller
statistical significance of short-baseline neutrino oscillations
of RSRF(N/R) ($2.6\sigma$) with respect to RSRF(N/DB) ($3.1\sigma$).
However,
one can see from Fig.~\ref{fig:see-gal-spe-sig}
that the $3\sigma$ allowed region of RSRF(N/R)
extends to large values of $\sin^2\!2\vartheta_{ee}$
for large values of $\Delta{m}^2_{41}$,
leading to a relative compatibility with the Gallium allowed regions.
On the other hand,
the $3\sigma$ allowed regions of RSRF(N/DB)
are closed and lie at values of $\sin^2\!2\vartheta_{ee}$
that are incompatible with the Gallium allowed regions.

Note that since the data sets considered in this paper are not the same as those of Ref.~\cite{Berryman:2021yan}, we obtain quantitatively different results.
As noted above,
the combined analysis of reactor spectral ratios in Ref.~\cite{Berryman:2021yan}
includes the Neutrino-4 data, which shift the best fit value into the region allowed by the analysis of the Gallium data (see Tab.~1 of Ref.~\cite{Berryman:2021yan}). As can be seen in the lower right panel of Fig.~2 of Ref.~\cite{Berryman:2021yan}, the 2$\sigma$ island around the best fit value lies at relatively large mixing and has a sizable overlap with the region preferred by the Gallium analysis, shown in the left panel of Fig.~4 of Ref.~\cite{Berryman:2021yan}. Therefore, the authors of Ref.~\cite{Berryman:2021yan} obtain a large parameter goodness of fit $p$-value for the combined analysis of the reactor spectral ratio data and Gallium data (see Tab.~5 of Ref.~\cite{Berryman:2021yan}). In our combined reactor spectral ratio analysis, using the newest DANSS data and omitting the Neutrino-4 data leads to regions which are quite small at the 2$\sigma$ level (even at the 3$\sigma$ level for RSRF(N/DB)) and far away from the Gallium allowed regions. Therefore, we find a significant tension between the fits of the reactor spectral ratio data and the Gallium data.

\begin{figure*}
\centering
\subfigure[]{ \label{fig:rates-ratios-neos17-2s}
\includegraphics[width=0.48\linewidth]{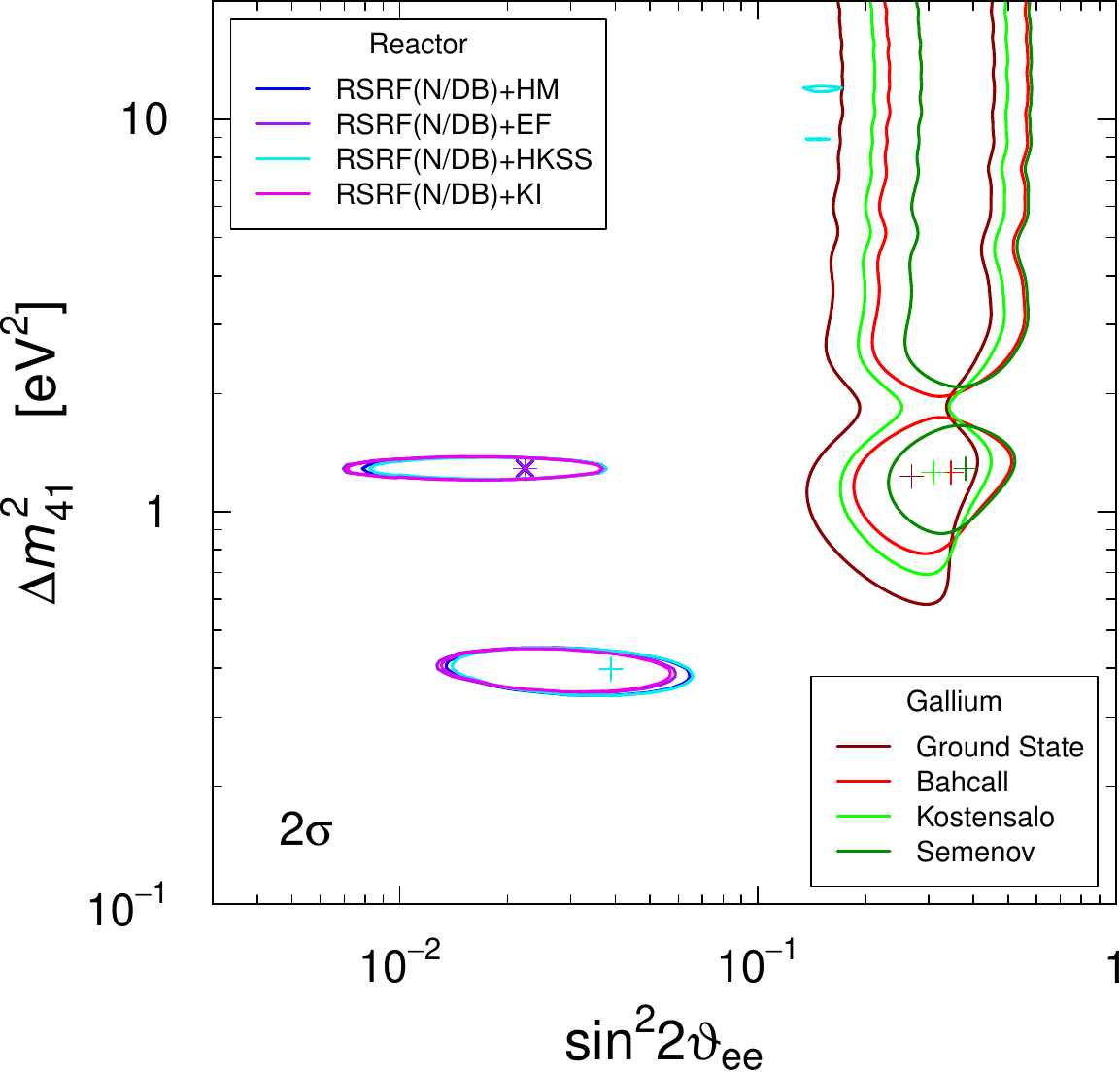}
}
\subfigure[]{ \label{fig:rates-ratios-neos21-2s}
\includegraphics[width=0.48\textwidth]{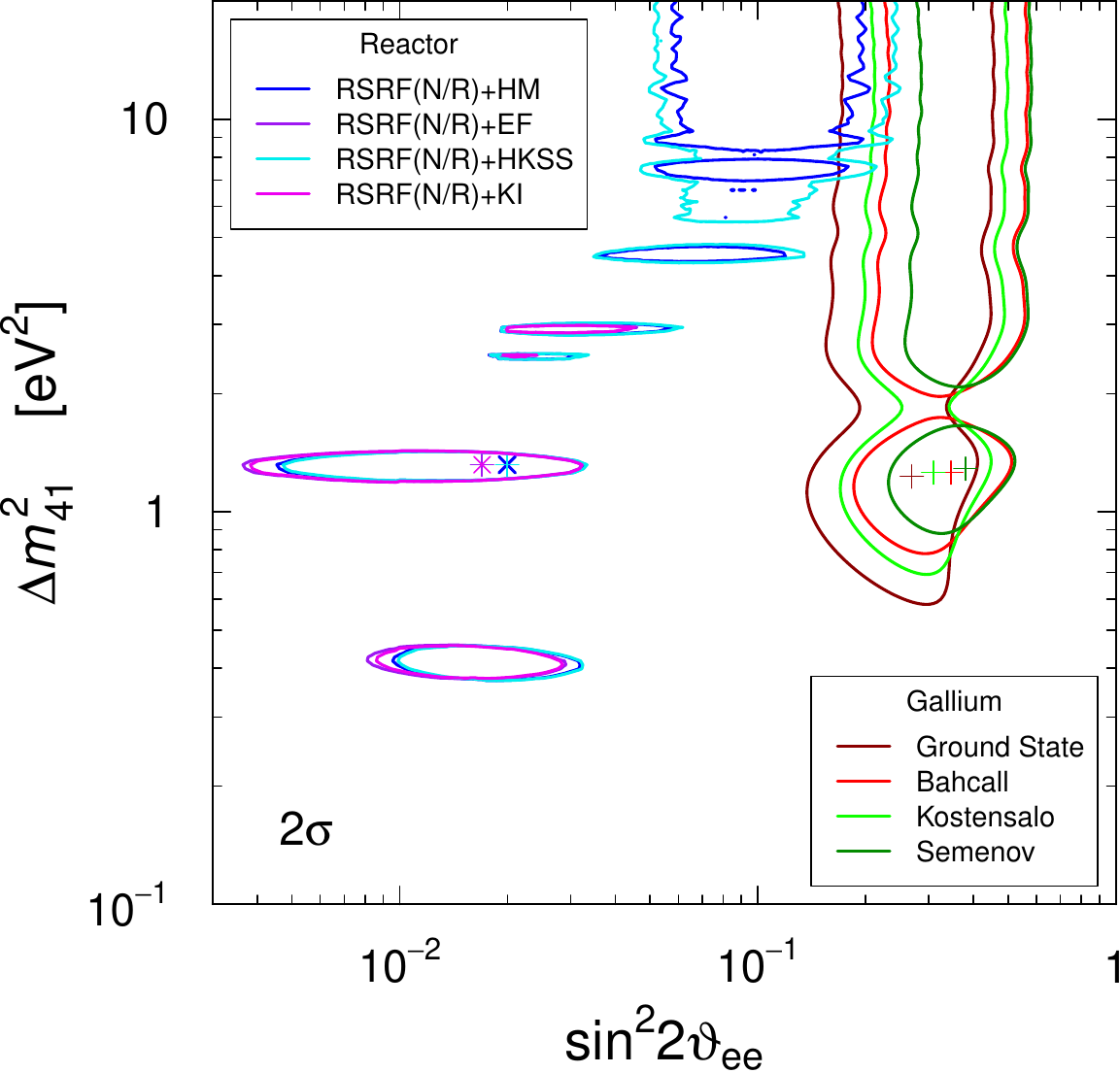}
}
\\
\subfigure[]{ \label{fig:rates-ratios-neos17-3s}
\includegraphics[width=0.48\linewidth]{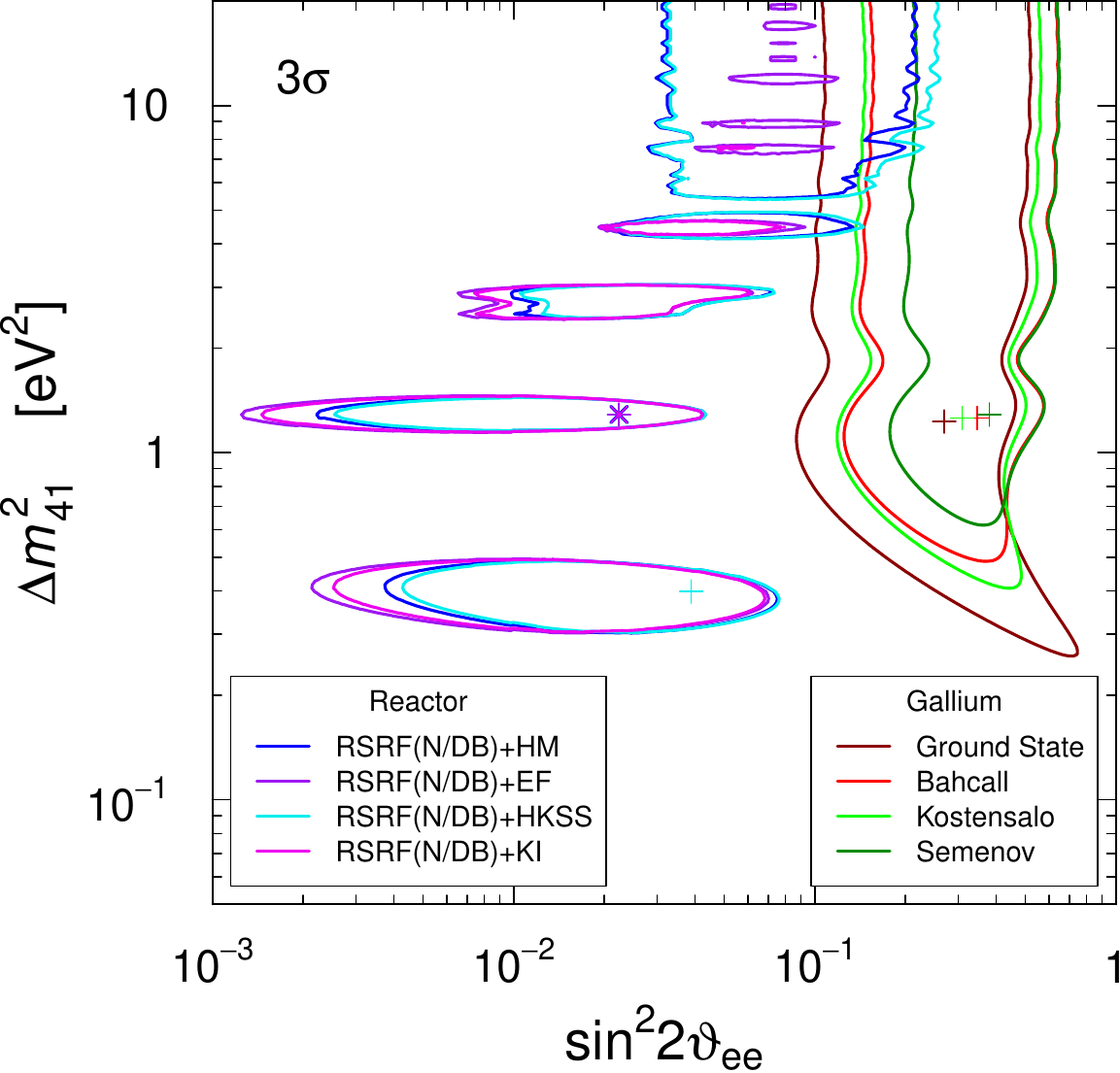}
}
\subfigure[]{ \label{fig:rates-ratios-neos21-3s}
\includegraphics[width=0.48\textwidth]{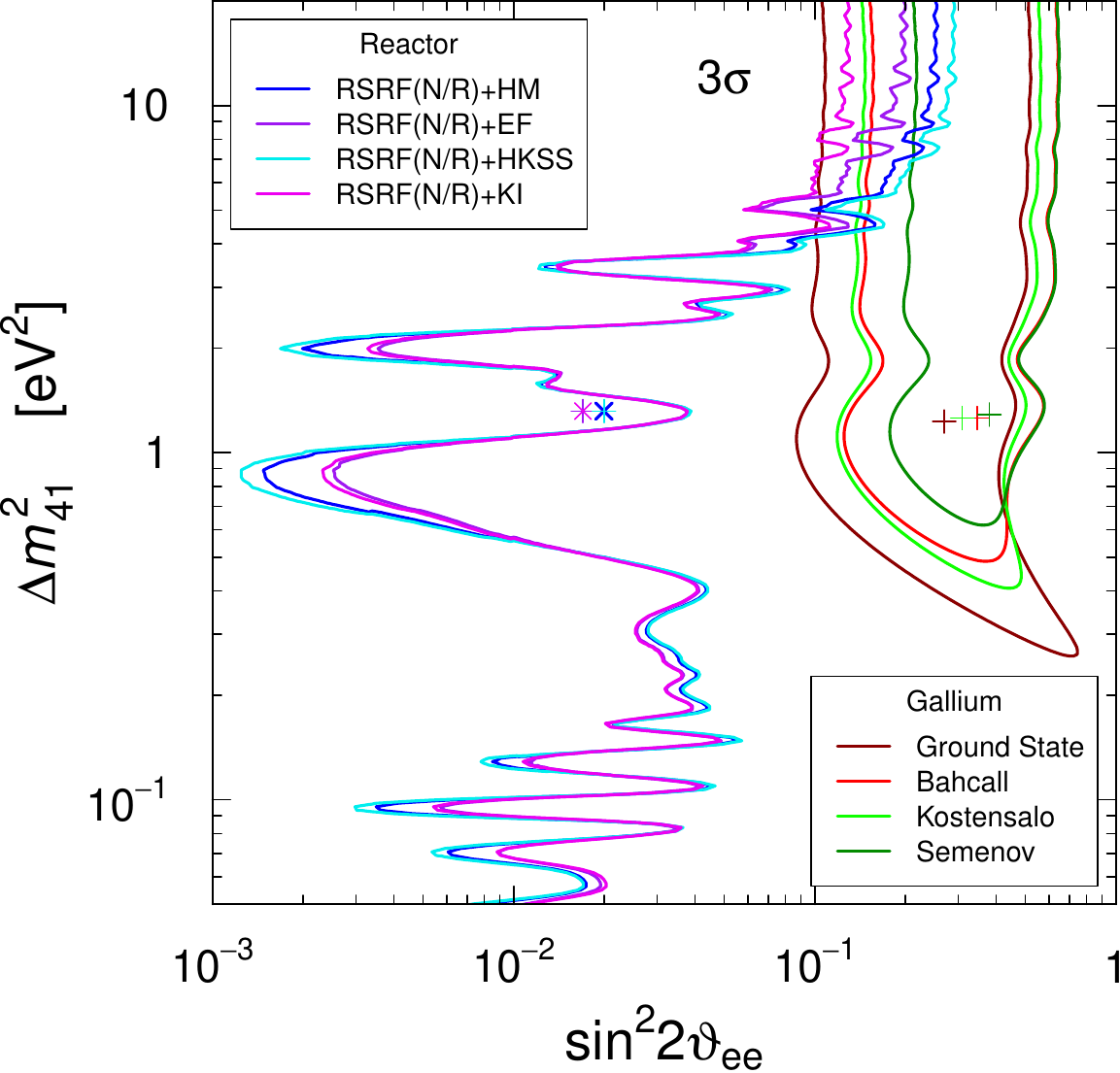}
}
\caption{\label{fig:rates-ratios}
Comparison of the contours delimiting the
[\subref{fig:rates-ratios-neos17-2s} and \subref{fig:rates-ratios-neos21-2s}]
$2\sigma$ and
[\subref{fig:rates-ratios-neos17-3s} and \subref{fig:rates-ratios-neos21-3s}]
$3\sigma$
allowed regions in the
($\sin^2\!2\vartheta_{ee},\Delta{m}^2_{41}$)
plane obtained from the combined analysis of the data of the
reactor rate experiments with different flux models
and
the spectral ratio experiments
with those obtained from the  Gallium data with different cross sections.
The figures differ by the use of
[\subref{fig:rates-ratios-neos17-2s} and \subref{fig:rates-ratios-neos17-3s}]
NEOS/Daya Bay~\cite{NEOS:2016wee}
or
[\subref{fig:rates-ratios-neos21-2s} and \subref{fig:rates-ratios-neos21-3s}]
NEOS/RENO~\cite{RENO:2020hva}
spectral ratio data.
The best-fit points are indicated by crosses.
}
\end{figure*}

\section{Combined short-baseline reactor spectral ratios and rates}
\label{sec:rates-ratios}

It is interesting to combine the results of the analyses
of the reactor spectral ratio data
with the results of the analyses
of the reactor rate data
discussed in Section~\ref{sec:rates}
and to compare the combination with the results of the
analyses of the Gallium data discussed in Section~\ref{sec:gallium}.
Fig.~\ref{fig:rates-ratios}
shows the comparison of the corresponding $2\sigma$ and $3\sigma$ allowed regions.

Comparing Fig.~\ref{fig:rates-ratios-neos17-2s}
with Fig.~\ref{fig:see-gal-spe-sig},
one can see that the addition of the reactor rate data
leads to a better localization of the reactor $2\sigma$ allowed regions
in the RSRF(N/DB) analysis, for all the reactor flux models.
In this case there are only two $2\sigma$-allowed islands at
$\Delta m_{41}^2 \simeq 1.3 \, \text{eV}^2$
and
$\Delta m_{41}^2 \simeq 0.4 \, \text{eV}^2$
for the HM, EF, and KI reactor flux models.
The first $2\sigma$-allowed island surrounds the best-fit point at
$\sin^2\!2\vartheta_{ee}=0.022$.
With the HKSS reactor flux models
there is also a third small $2\sigma$-allowed island at
$\Delta m_{41}^2 \simeq 12 \, \text{eV}^2$
and
$\sin^2\!2\vartheta_{ee}=0.13-0.17$,
indicating a larger compatibility with the Gallium data.
However, in this case the best-fit point
lies in the second $2\sigma$-allowed island at
$\sin^2\!2\vartheta_{ee}=0.04$.
Figure~\ref{fig:rates-ratios-neos17-3s}
shows the $3\sigma$-allowed regions
obtained with the combined analysis of RSRF(N/DB) and reactor rates.
One can see that the HM and HKSS reactor flux models
give large $3\sigma$-allowed regions at large values of $\Delta m_{41}^2$,
which include relatively large values of
$\sin^2\!2\vartheta_{ee}$
and partially overlap
with the Gallium $3\sigma$-allowed regions.
Also the EF reactor flux model has some small $3\sigma$-allowed regions
at large values of $\Delta m_{41}^2$, but
there is only little overlap
with the Gallium $3\sigma$-allowed regions and only when considering the extreme Ground State cross section model.
The KI reactor flux model does not yield any $3\sigma$-allowed region
at large values of $\Delta m_{41}^2$
and is in strongest tension with the Gallium $3\sigma$-allowed region.
The tension in the different cases is quantified by the values of the
parameter goodness of fit
in Tab.~\ref{tab:rea-gal-pgf},
where one can see that the fits with all the
Gallium cross section models
are in very strong and unacceptable tension with the
reactor RSRF(N/DB) + KI fit.
The tension is smaller, but still very strong and unacceptable
in the case of the reactor RSRF(N/DB) + EF fit.
The $\text{GoF}_{\text{PG}}$ is below 0.1\% for the  RSRF(N/DB) + HM fit
and also below 1\% for the RSRF(N/DB) + HKSS fit.

\begin{table*}
\centering
\begin{tabular}{|c|cc|cc|cc|cc|}
\cline{2-9}
\multicolumn{1}{c|}{}
&
\multicolumn{8}{c|}{RSRF(N/DB) + Reactor Rates}
\\
\cline{2-9}
\multicolumn{1}{c|}{}
&
\multicolumn{2}{c|}{HM}
&
\multicolumn{2}{c|}{HKSS}
&
\multicolumn{2}{c|}{EF}
&
\multicolumn{2}{c|}{KI}
\\
\multicolumn{1}{c|}{}
&
$\Delta\chi^{2}_{\text{PG}}$
&
$\text{GoF}_{\text{PG}}$
&
$\Delta\chi^{2}_{\text{PG}}$
&
$\text{GoF}_{\text{PG}}$
&
$\Delta\chi^{2}_{\text{PG}}$
&
$\text{GoF}_{\text{PG}}$
&
$\Delta\chi^{2}_{\text{PG}}$
&
$\text{GoF}_{\text{PG}}$
\\
\hline
Ground State
&
$14.30$
&
$0.078\%$
&
$11.36$
&
$0.34\%$
&
$19.57$
&
$0.0056\%$
&
$21.81$
&
$0.0018\%$
\\
\hline
Bahcall
&
$18.33$
&
$0.01\%$
&
$15.16$
&
$0.051\%$
&
$23.60$
&
$0.00075\%$
&
$26.02$
&
$0.00022\%$
\\
\hline
Kostensalo
&
$17.04$
&
$0.02\%$
&
$13.80$
&
$0.1\%$
&
$22.30$
&
$0.0014\%$
&
$27.51$
&
$0.00011\%$
\\
\hline
Semenov
&
$23.22$
&
$0.00091\%$
&
$19.39$
&
$0.0061\%$
&
$28.28$
&
$0.000072\%$
&
$36.85$
&
$0.00000099\%$
\\
\hline
\multicolumn{1}{c|}{}
&
\multicolumn{8}{c|}{RSRF(N/R) + Reactor Rates}
\\
\cline{2-9}
\multicolumn{1}{c|}{}
&
\multicolumn{2}{c|}{HM}
&
\multicolumn{2}{c|}{HKSS}
&
\multicolumn{2}{c|}{EF}
&
\multicolumn{2}{c|}{KI}
\\
\multicolumn{1}{c|}{}
&
$\Delta\chi^{2}_{\text{PG}}$
&
$\text{GoF}_{\text{PG}}$
&
$\Delta\chi^{2}_{\text{PG}}$
&
$\text{GoF}_{\text{PG}}$
&
$\Delta\chi^{2}_{\text{PG}}$
&
$\text{GoF}_{\text{PG}}$
&
$\Delta\chi^{2}_{\text{PG}}$
&
$\text{GoF}_{\text{PG}}$
\\
\hline
Ground State
&
$10.12$
&
$0.63\%$
&
$6.94$
&
$3.1\%$
&
$15.59$
&
$0.041\%$
&
$21.04$
&
$0.0027\%$
\\
\hline
Bahcall
&
$14.14$
&
$0.085\%$
&
$10.72$
&
$0.47\%$
&
$19.61$
&
$0.0055\%$
&
$25.63$
&
$0.00027\%$
\\
\hline
Kostensalo
&
$12.84$
&
$0.16\%$
&
$9.36$
&
$0.93\%$
&
$18.30$
&
$0.011\%$
&
$24.89$
&
$0.00039\%$
\\
\hline
Semenov
&
$19.04$
&
$0.0073\%$
&
$15.00$
&
$0.055\%$
&
$24.29$
&
$0.00053\%$
&
$32.99$
&
$0.0000068\%$
\\
\hline
\end{tabular}

\caption{ \label{tab:rea-gal-pgf}
$\chi^2$ difference
$\Delta\chi^{2}_{\text{PG}}$
of the parameter goodness of fit test~\cite{Maltoni:2003cu}
applied to the comparison of the neutrino oscillation fits of the reactor rates and spectral ratio data
with the Gallium data
using the four Gallium detection cross sections
in Tab.~\ref{tab:gallium_cross_sections}.
The values of the corresponding parameter goodness of fit
$\text{GoF}_{\text{PG}}$
are calculated
with two degrees of freedom corresponding to the two
common oscillation parameters
$\sin^2\!2\vartheta_{ee}$ and $\Delta{m}^2_{41}$.
The column titles RSRF(N/DB) and RSRF(N/R)
indicate, respectively,
the combined fits of NEOS/Daya Bay and NEOS/RENO data with those
of the other reactor spectral ratio experiments discussed in the text
(DANSS, PROSPECT, STEREO, and Bugey-3).
The column titles HM, HKSS, EF, and KI
refer to the four reactor neutrino fluxes discussed in Section~\ref{sec:rates}.
}
\end{table*}

Let us now discuss the results
of the combined analysis of RSRF(N/R) and the reactor rates,
that are shown in Figs.~\ref{fig:rates-ratios-neos21-2s}
and~\ref{fig:rates-ratios-neos21-3s}.
Comparing with Figs.~\ref{fig:rates-ratios-neos17-2s}
and~\ref{fig:rates-ratios-neos17-3s},
one can see that the constraints on the mixing parameters are looser
than in the combined analysis of RSRF(N/DB) and reactor rates.
In particular,
from Fig.~\ref{fig:rates-ratios-neos21-2s}
one can see that
the combined analyses of RSRF(N/R) and reactor rates
with the HM and HKSS reactor flux models
yields large $2\sigma$-allowed regions
at large values of $\Delta m_{41}^2$
and relatively large values of
$\sin^2\!2\vartheta_{ee}$,
with partial overlaps with the
Gallium $2\sigma$-allowed regions
obtained with the Ground State, Bahcall, and Kostensalo
Gallium cross section models.
Figure~\ref{fig:rates-ratios-neos21-3s}
shows that the combined analyses of RSRF(N/R) and reactor rates
with all the reactor flux models
yield only exclusion curves at $3\sigma$,
but there are significant overlaps of the allowed regions
with some of the Gallium $3\sigma$-allowed regions.
Therefore,
as shown in Tab.~\ref{tab:rea-gal-pgf},
the tension between reactor and Gallium data
is smaller for
the combined analyses of RSRF(N/R) and reactor rates
than for
the combined analyses of RSRF(N/DB) and reactor rates.
One can see from Tab.~\ref{tab:rea-gal-pgf}
that nevertheless the tension is very strong
and unacceptable
in the case of the
reactor RSRF(N/R) + KI fit.
The tension is smaller, but still with $\text{GoF}_{\text{PG}}$ well below 0.1\%
in the case of the reactor RSRF(N/R) + EF fit.
The $\text{GoF}_{\text{PG}}$ is also below 1\% for the  RSRF(N/R) + HM fit.
Only when considering the RSRF(N/R) + HKSS fit
and the Ground State Gallium cross section model we find $\text{GoF}_{\text{PG}}>1\%$. 

%%%%%%%%%%%%%%%%%%%%%%%%%%%%%%%%%%%%%%%%%%%%%%%%%%%%%%%%

\section{The KATRIN limit}
\label{sec:KATRIN}

The Karlsruhe Tritium Neutrino (KATRIN) experiment provides the current strongest laboratory limit on the absolute mass of light neutrinos. It aims to measure the end point region of the spectrum of molecular tritium $\beta$-decay ($\text{T}_2\rightarrow {}^3\text{HeT}^+ + e^- + \bar\nu_e$) with high enough precision to discern the shape distortion caused by tiny neutrino masses. The first (second) data taking campaign set an upper bound of 1.1~eV (0.9~eV) at 90\% C.L. on the effective electron neutrino mass $m_{\beta}$, which, in the standard three-neutrino mixing framework is defined as
\begin{equation}
m_{\beta}^2
=
\sum_{i=1}^{3} |U_{ei}|^2 m_i^2
.
\label{m_beta}
\end{equation}
An upper bound of 0.8~eV at 90\% C.L. is obtained when data from both campaigns are combined~\cite{KATRIN:2021uub}. Data from KATRIN has already been used to search for light sterile neutrinos, see Refs.~\cite{Giunti:2019fcj,KATRIN:2020dpx,KATRIN:2022ith}.
In this work we use the data from the second campaign to constrain $\Delta m^2_{41}$ and $\sin^22\vartheta_{ee}$. While we follow the general analysis approach of the KATRIN Collaboration, as described in Ref.~\cite{KATRIN:2022ith}, there are some differences which we discuss below.

The differential spectrum $R_{\beta}(E)$ of $\beta$-decay can be calculated using Fermi's Golden Rule:
\begin{multline}
    R_{\beta}(E,m_{\beta}^2)=\frac{G_F^2\,\textrm{cos}^2\theta_C}{2\pi^3}|M_{\textrm{nuc}}|^2 F(E,Z)\\\times(E+m_e)\sqrt{(E+m_e)^2-m_e^2}\\ \times \sum_j P_j \epsilon_j \sqrt{\epsilon_j^2-m_{\beta}^2}\,\Theta(\epsilon_j-m_{\beta}^2)
    .
\end{multline}
Here $G_{F}$ is the Fermi constant, $\theta_C$ is the Cabibbo angle, $M_{\textrm{nuc}}$ is the nuclear matrix element, $m_e$ is the electron mass, $E$ is the kinetic energy of the emitted electron, $F(E, Z)$ is the relativistic Fermi function, $\epsilon_j=E_0-E-V_j$ are the neutrino energies, where $E_0$ is the end point energy of gaseous molecular tritium ($T_2$) and $V_j$ are the energies of the various excited final states which occur with probabilities $P_j$~\cite{Saenz:2000dul}. 
Measurements of the atomic mass difference of tritium and ${}^3$He~\cite{Myers:2015lca} have found the $Q$-value to be $18575.72 \pm 0.07~\text{eV}$. This $Q$-value corresponds to the end point $E_0 = 18574.21 \pm 0.6~\text{eV}$~\cite{KATRIN:2021uub}. 

The KATRIN detector measures the integral spectrum which, at a retarding potential $qU_i$, is given by
\begin{multline}
        R_{\textrm{model}}(qU_i)=A_{\textrm{sig}}N_T\int_{qU_i}^{E_0} R_{\beta}(E)f(E-qU_i)dE\\ + R_{\textrm{bg}}(qU_i)\,.
\end{multline}
The differential spectrum $R_{\beta}(E)$ is convoluted with the experimental response function $f(E-qU_i)$. The response function describes the probability for an electron emitted with kinetic energy $E$ to reach the detector when the retarding potential is $qU_i$. We use the green dotted line in Fig.~5 of Ref.~\cite{KATRIN:2021uub} for the response function. In the above equation, $N_T$ and $A_{sig}$ are the effective number of tritium atoms in the source and the signal amplitude respectively. Since we are interested in performing a shape only analysis, $N_T$ and $A_{sig}$ are combined into a normalization factor $N$ which is left free in the analysis. The background rate $R_{bg}(qU_i)$ has three components
\begin{equation}
    R_{\textrm{bg}}(qU_i)=R^{\textrm{\,base}}_{\textrm{bg}}+R^{\,qU}_{\textrm{bg}}(qU_i)+R^{\textrm{\,Penning}}_{\textrm{bg}}(t(qU_i)),
\end{equation}
where $t(qU_i)$ is the time spent at the retarding potential $qU_i$. The dominant contribution comes from the flat constant rate $R^{\textrm{base}}_{\textrm{bg}}$. A hypothetical $qU$-dependent background $R^{qU}_{\textrm{bg}}(qU_i)$, with slope constrained to $(0.0\pm4.74)$\,mcps/keV, and a scan-time dependent background $R^{\textrm{Penning}}_{\textrm{bg}}(t(qU_i))$, constrained to $(3\pm3)\,\mu$cps/s, give smaller contributions. Pull terms are added to the $\chi^2$ function to account for the constraints on the sub-dominant background rates. 

For the case of a sterile neutrino with mass $m_4$ and active-sterile mixing $|U_{e4}|$, the expected rate is
\begin{multline}
    R_{\textrm{pred}}(E,m_{\beta}^2,m^2_4,|U_{e4}|^2)=(1-|U_{e4}|^2)\,R_{\beta}(E,m_{\beta}^2)\\+|U_{e4}|^2\,R_{\beta}(E,m_4^2)\,.
\end{multline}
To infer the parameters of interest, we perform fits with fixed $\Delta m^2_{41}$ and $|U_{e4}|^2$ pairs, minimizing the function
\begin{multline}
    \chi^2=[\Delta\Vec{R}]^{T} C^{-1}[\Delta\Vec{R}]\\+\left(\frac{m^{qU}}{4.74}\right)^2\,+\,\left(\frac{m^{\textrm{Penn}}-3}{3}\right)^2\,+\,\left(\frac{E_0-18574.21}{0.6}\right)^2,
\end{multline}
where $\Delta\Vec{R}=\Vec{R}_{\textrm{obs}}(\Vec{qU})-\Vec{R}_{\textrm{pred}}(\Vec{qU},\Vec{\eta})$, $m^{qU}$ and $m^{\textrm{Penn}}$ are the slopes of the background components and $C^{-1}$ is the (inverse) covariance matrix provided by the collaboration. To verify our analysis method we first reproduced the standard three-neutrino results and obtained an upper bound of $m_{\beta}<0.83$~eV at 90\% C.L. and a best fit $m_{\beta}^2=0.1\,\textrm{eV}^2$. These results validate our analysis and so we proceed with the sterile analysis.

\begin{figure}[h]
\centering
\includegraphics[width=\linewidth]{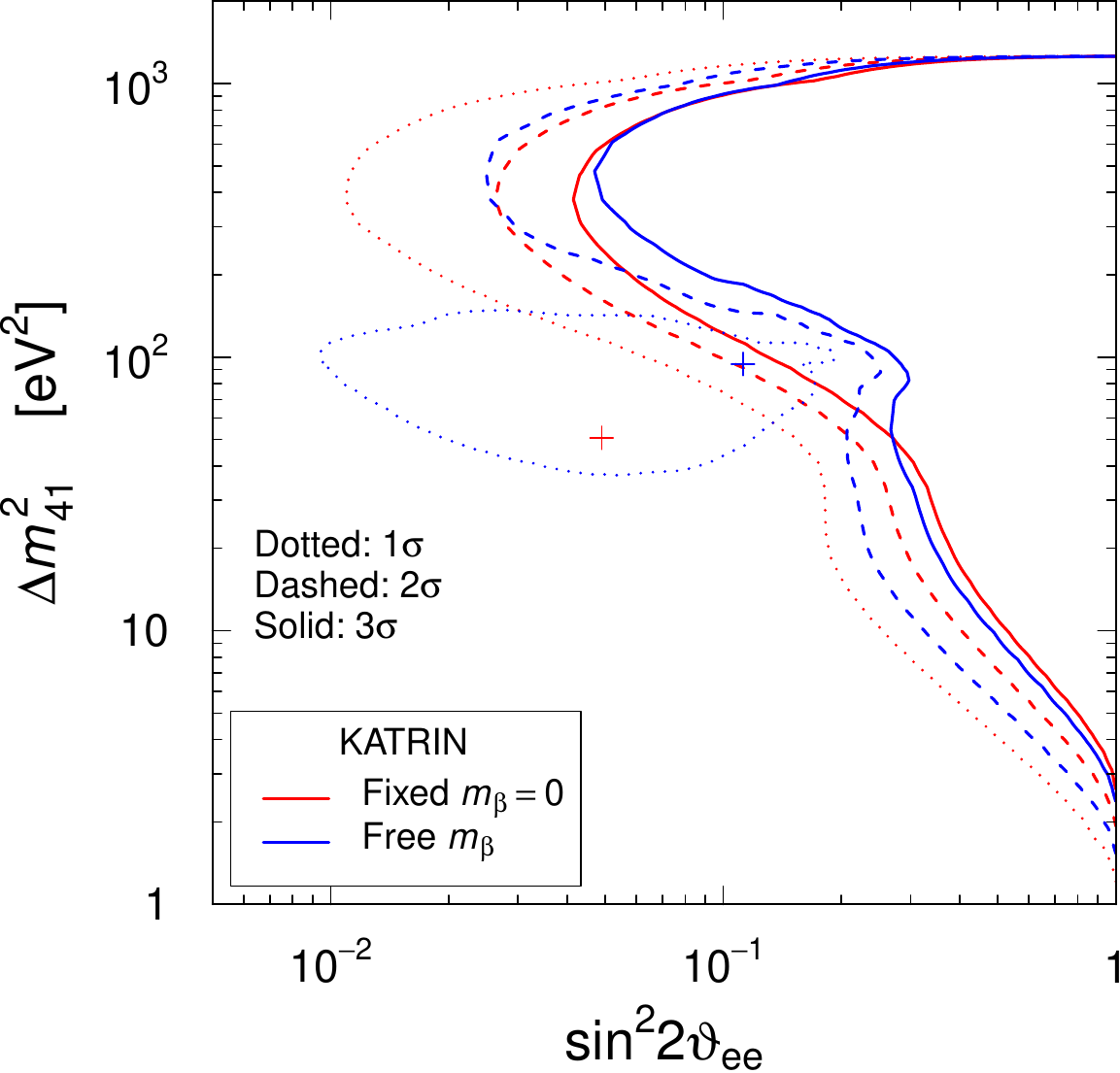}
\caption{\label{fig:KAT-sterile}
Contours delimiting the 1$\sigma$, 2$\sigma$, 3$\sigma$
allowed regions in the
($\sin^2\!2\vartheta_{ee},\Delta{m}^2_{41}$)
plane
obtained from the analysis of KATRIN run 2 data.
Red curves correspond to the analysis with fixed $m_{\beta}=0$,
while the blue curves are obtained after marginalizing over a free $m_{\beta}$.
The best-fit points are indicated by crosses.
}
\end{figure}

\begin{figure*}
\centering
\subfigure[]{ \label{fig:rates-ratios-tritium-neos17-2s}
\includegraphics[width=0.48\linewidth]{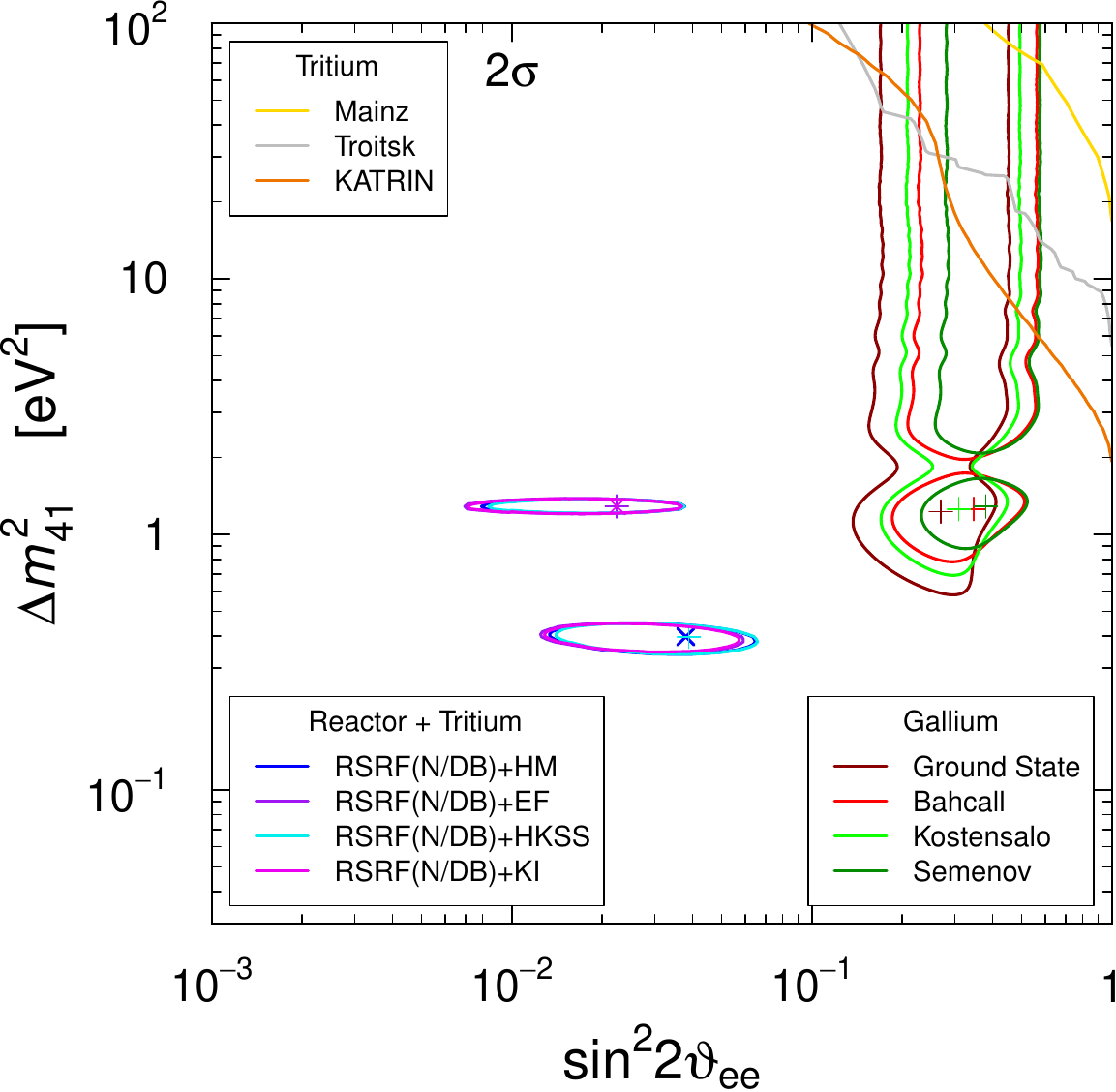}
}
\subfigure[]{ \label{fig:rates-ratios-tritium-neos21-2s}
\includegraphics[width=0.48\textwidth]{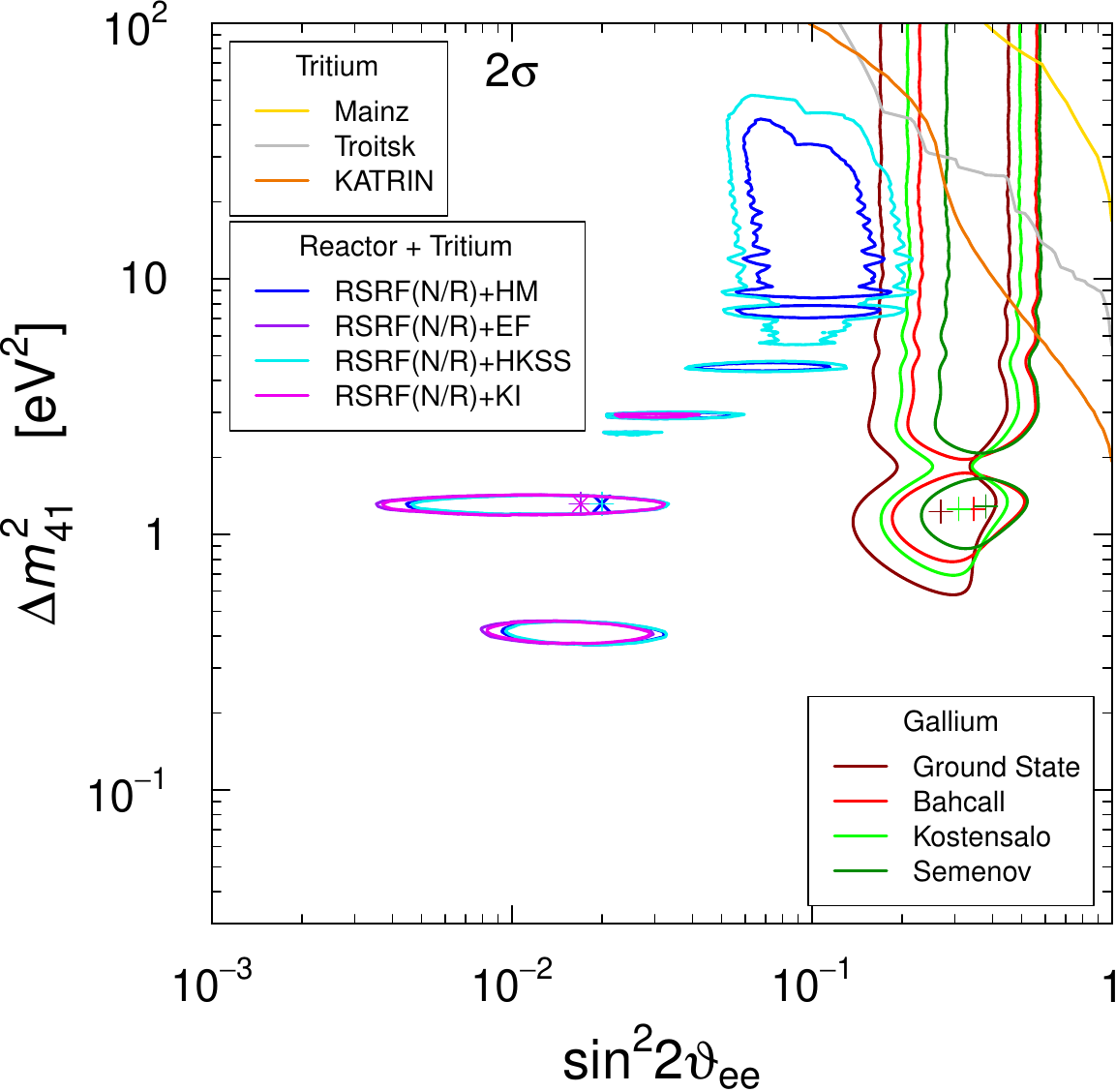}
}
\\
\subfigure[]{ \label{fig:rates-ratios-tritium-neos17-3s}
\includegraphics[width=0.48\linewidth]{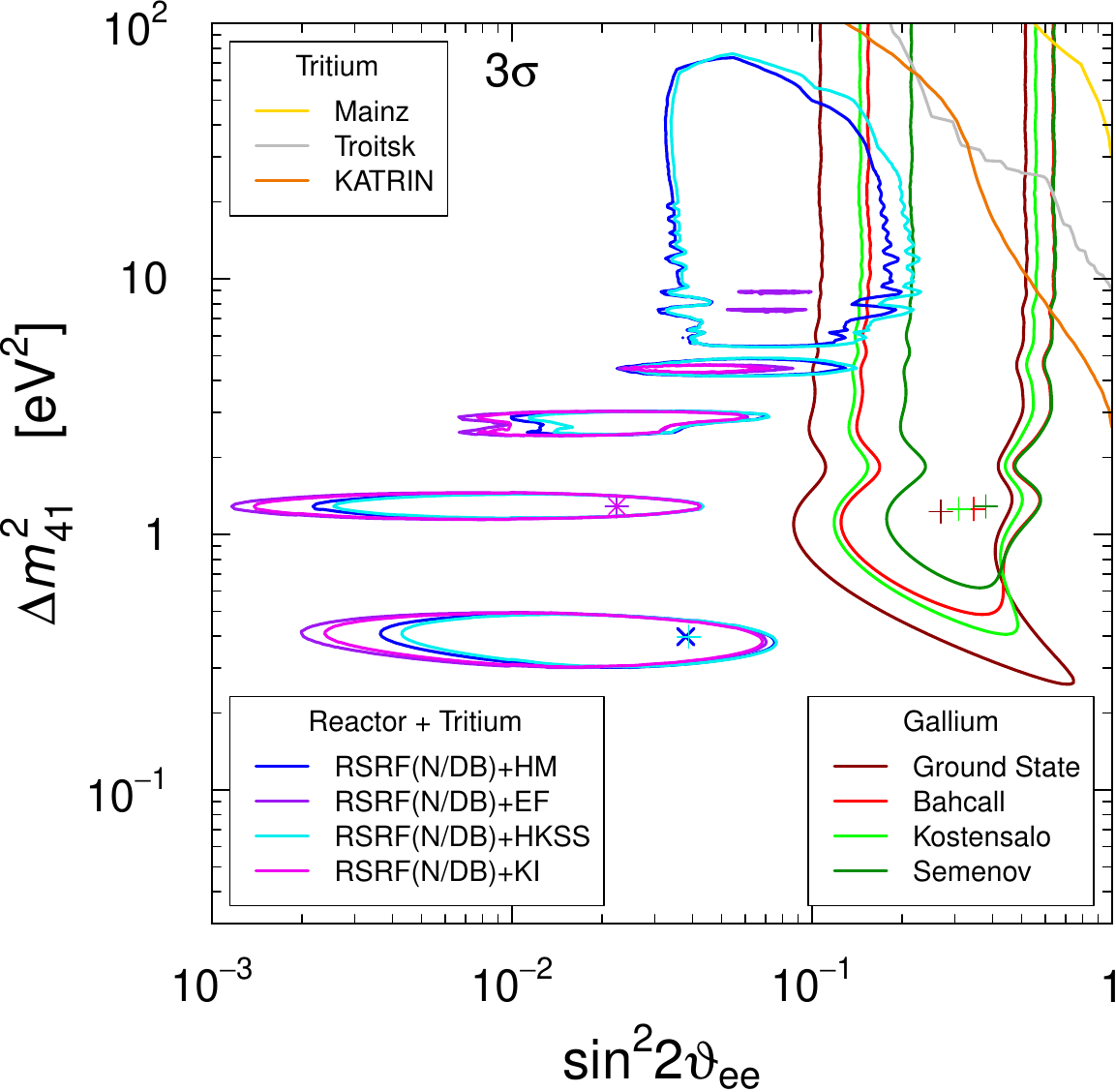}
}
\subfigure[]{ \label{fig:rates-ratios-tritium-neos21-3s}
\includegraphics[width=0.48\textwidth]{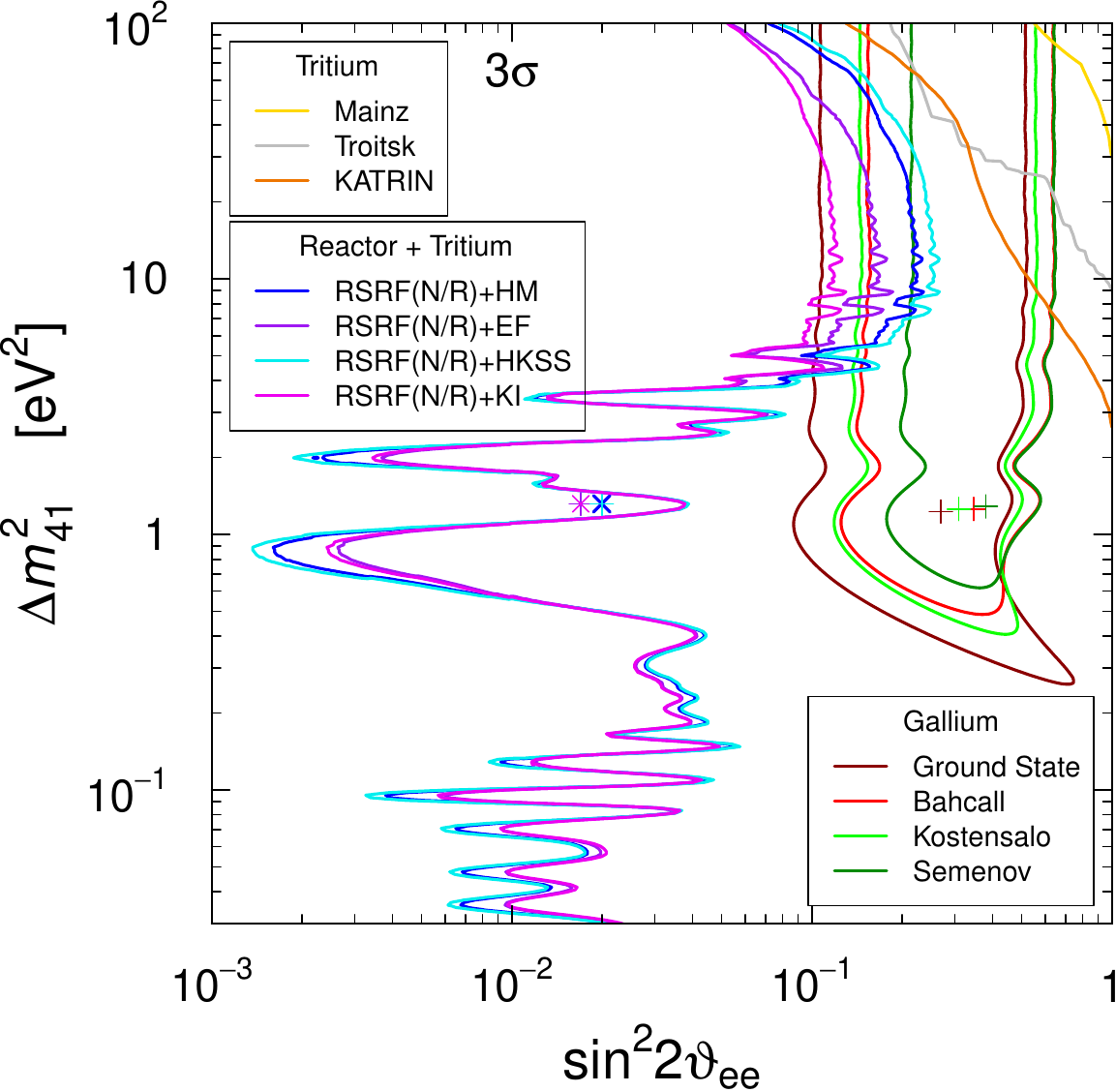}
}
\caption{\label{fig:rates-ratios-tritium}
Comparison of the contours delimiting the
[\subref{fig:rates-ratios-tritium-neos17-2s} and \subref{fig:rates-ratios-tritium-neos21-2s}]
$2\sigma$ and
[\subref{fig:rates-ratios-tritium-neos17-3s} and \subref{fig:rates-ratios-tritium-neos21-3s}]
$3\sigma$
allowed regions in the
($\sin^2\!2\vartheta_{ee},\Delta{m}^2_{41}$)
plane obtained from the combined analysis of the data of the
reactor rate experiments with different flux models,
the spectral ratio experiments,
and the Tritium experiments
with those obtained from the  Gallium data with different cross sections.
The figures differ by the use of
[\subref{fig:rates-ratios-tritium-neos17-2s} and \subref{fig:rates-ratios-tritium-neos17-3s}]
NEOS/Daya Bay~\cite{NEOS:2016wee}
or
[\subref{fig:rates-ratios-tritium-neos21-2s} and \subref{fig:rates-ratios-tritium-neos21-3s}]
NEOS/RENO~\cite{RENO:2020hva}
spectral ratio data.
The best-fit points are indicated by crosses.
}
\end{figure*}

\begin{figure}
\centering
\includegraphics[width=\linewidth]{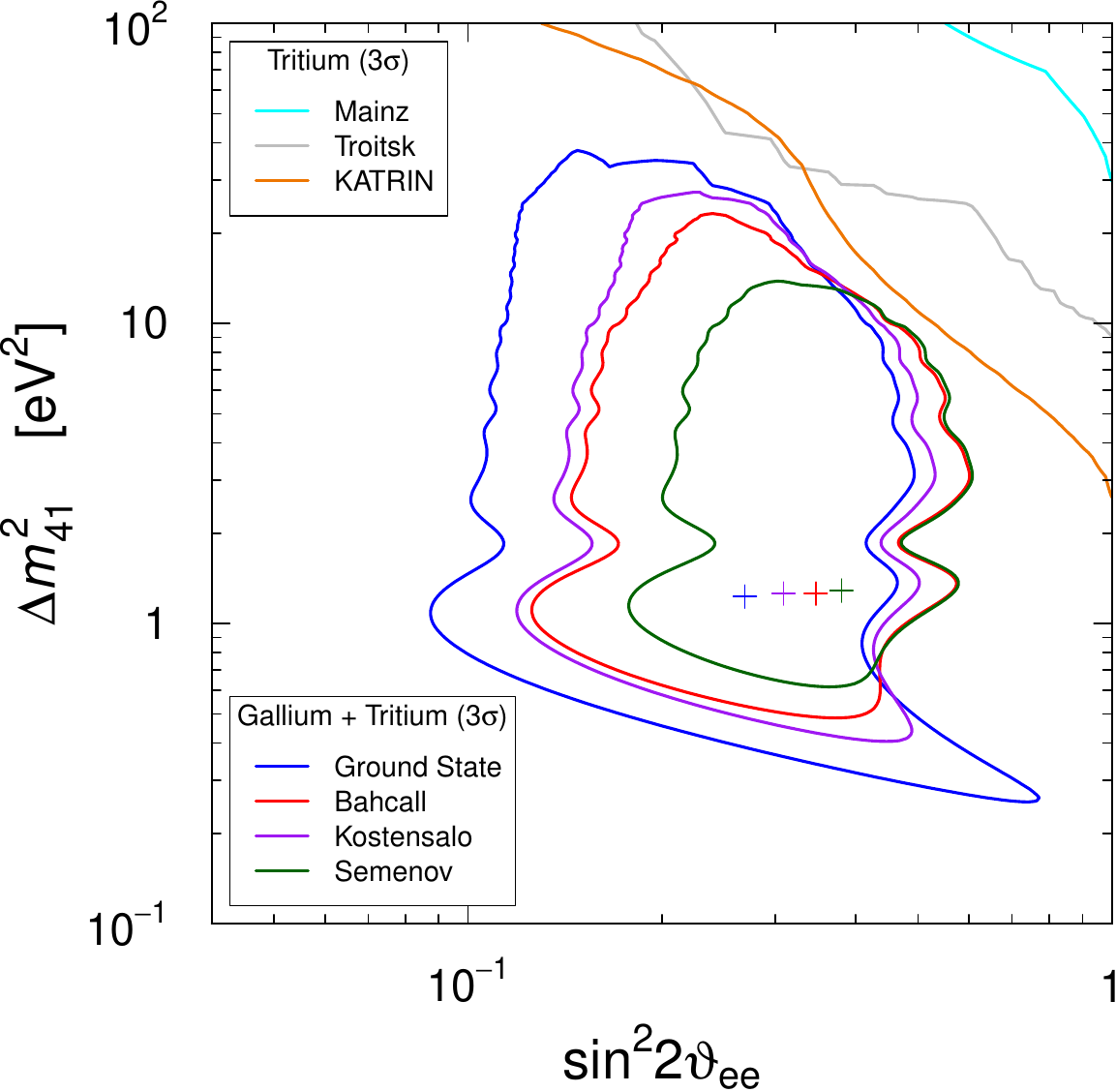}
\caption{\label{fig:see-gal-tri-3s}
Contours delimiting the $3\sigma$ allowed regions in the
($\sin^2\!2\vartheta_{ee},\Delta{m}^2_{41}$)
plane obtained from the combined analysis
of the Gallium data with different cross sections
and the Tritium data.
The figure shows also the $3\sigma$ exclusion curves
of the Mainz, Troitsk, and KATRIN
Tritium experiments.
}
\end{figure}

\begin{table*}
\centering
\begin{tabular}{|c|cc|cc|cc|cc|}
\cline{2-9}
\multicolumn{1}{c|}{}
&
\multicolumn{8}{c|}{RSRF(N/DB) + Reactor Rates + Tritium}
\\
\cline{2-9}
\multicolumn{1}{c|}{}
&
\multicolumn{2}{c|}{HM}
&
\multicolumn{2}{c|}{HKSS}
&
\multicolumn{2}{c|}{EF}
&
\multicolumn{2}{c|}{KI}
\\
\multicolumn{1}{c|}{}
&
$\Delta\chi^{2}_{\text{PG}}$
&
$\text{GoF}_{\text{PG}}$
&
$\Delta\chi^{2}_{\text{PG}}$
&
$\text{GoF}_{\text{PG}}$
&
$\Delta\chi^{2}_{\text{PG}}$
&
$\text{GoF}_{\text{PG}}$
&
$\Delta\chi^{2}_{\text{PG}}$
&
$\text{GoF}_{\text{PG}}$
\\
\hline
Ground State
&
$15.69$
&
$0.039\%$
&
$13.17$
&
$0.14\%$
&
$20.82$
&
$0.003\%$
&
$21.82$
&
$0.0018\%$
\\
\hline
Bahcall
&
$19.86$
&
$0.0049\%$
&
$17.19$
&
$0.019\%$
&
$25.06$
&
$0.00036\%$
&
$26.03$
&
$0.00022\%$
\\
\hline
Kostensalo
&
$18.63$
&
$0.009\%$
&
$15.87$
&
$0.036\%$
&
$23.83$
&
$0.00067\%$
&
$27.52$
&
$0.00011\%$
\\
\hline
Semenov
&
$25.22$
&
$0.00033\%$
&
$21.94$
&
$0.0017\%$
&
$30.42$
&
$0.000025\%$
&
$37.42$
&
$0.00000075\%$
\\
\hline
\multicolumn{1}{c|}{}
&
\multicolumn{8}{c|}{RSRF(N/R) + Reactor Rates + Tritium}
\\
\cline{2-9}
\multicolumn{1}{c|}{}
&
\multicolumn{2}{c|}{HM}
&
\multicolumn{2}{c|}{HKSS}
&
\multicolumn{2}{c|}{EF}
&
\multicolumn{2}{c|}{KI}
\\
\multicolumn{1}{c|}{}
&
$\Delta\chi^{2}_{\text{PG}}$
&
$\text{GoF}_{\text{PG}}$
&
$\Delta\chi^{2}_{\text{PG}}$
&
$\text{GoF}_{\text{PG}}$
&
$\Delta\chi^{2}_{\text{PG}}$
&
$\text{GoF}_{\text{PG}}$
&
$\Delta\chi^{2}_{\text{PG}}$
&
$\text{GoF}_{\text{PG}}$
\\
\hline
Ground State
&
$11.56$
&
$0.31\%$
&
$8.72$
&
$1.3\%$
&
$16.96$
&
$0.021\%$
&
$21.49$
&
$0.0022\%$
\\
\hline
Bahcall
&
$15.76$
&
$0.038\%$
&
$12.74$
&
$0.17\%$
&
$21.19$
&
$0.0025\%$
&
$26.08$
&
$0.00022\%$
\\
\hline
Kostensalo
&
$14.49$
&
$0.071\%$
&
$11.40$
&
$0.33\%$
&
$19.97$
&
$0.0046\%$
&
$25.37$
&
$0.00031\%$
\\
\hline
Semenov
&
$21.04$
&
$0.0027\%$
&
$17.45$
&
$0.016\%$
&
$26.45$
&
$0.00018\%$
&
$33.56$
&
$0.0000052\%$
\\
\hline
\end{tabular}

\caption{ \label{tab:rea-tri-gal-pgf}
$\chi^2$ difference
$\Delta\chi^{2}_{\text{PG}}$
of the parameter goodness of fit test~\cite{Maltoni:2003cu}
applied to the comparison of the neutrino oscillation fits of the reactor rates, spectral ratio data, and tritium data
with the Gallium data
using the four Gallium detection cross sections
in Tab.~\ref{tab:gallium_cross_sections}.
The values of the corresponding parameter goodness of fit
$\text{GoF}_{\text{PG}}$
are calculated
with two degrees of freedom corresponding to the two
common oscillation parameters
$\sin^2\!2\vartheta_{ee}$ and $\Delta{m}^2_{41}$.
The column titles RSRF(N/DB) and RSRF(N/R)
indicate, respectively,
the combined fits of NEOS/Daya Bay and NEOS/RENO data with those
of the other reactor spectral ratio experiments discussed in the text
(DANSS, PROSPECT, STEREO, and Bugey-3).
The column titles HM, HKSS, EF, and KI
refer to the four reactor neutrino fluxes discussed in Section~\ref{sec:rates}.
}
\end{table*}

There are six nuisance parameters, the overall normalization $N$, the end point $E_0$, the three background parameters and the effective neutrino mass-squared $m^2_{\beta}$, which are marginalized over. We performed two analyses.
In one we fixed $m_{\beta}=0$ (so that $\Delta m^2_{41}=m^2_4$) and in the other $m_{\beta}$ was left free and marginalized over. 
The contours resulting from these analyzes are shown in Fig.~\ref{fig:KAT-sterile}. The red exclusion curves correspond to the analysis where we fix $m_\beta = 0$. The results of our analysis are in excellent agreement with the results obtained by the KATRIN collaboration~\cite{KATRIN:2022ith} in this case. We find that KATRIN data alone disfavor the preferred region from Gallium experiments for mass splittings $\Delta m_{41}^2\gtrsim 100$~eV$^2$. The global minimum of the $\chi^2$ function is $\chi^2=27.59$ at $\Delta m^2_{41}=50.6\,\textrm{eV}^2$ and $\sin^2\!2\vartheta_{ee}=0.05$ ($|U_{e4}|^2=0.013$). The significance over the null hypothesis is $\Delta\chi^2=0.38$. 
The blue contours in Fig.~\ref{fig:KAT-sterile} are obtained when marginalizing over the free $m_\beta$. In contrast to the collaboration's approach, we require that the neutrino mass is real ($m_\beta^2\geq 0$) and $m_4 > m_{\beta}$. 
The analysis with free $m_{\beta}$ gives minimum $\chi^2=25.27$ at $\Delta m^2_{41}=87.8\,\textrm{eV}^2\,(m^2_4=88.7\,\textrm{eV}^2,m_{\beta}^2=0.9\,\textrm{eV}^2)$ and $\textrm{sin}^2(\vartheta_{ee})=0.11$ ($|U_{e4}|^2=0.029$). The preference over the no sterile hypothesis is larger ($\Delta\chi^2=2.7$) but still not very significant. This explains the closed 1$\sigma$ contour in Fig.~\ref{fig:KAT-sterile}. Such a small preference has already been observed in former analyses~\cite{Giunti:2019fcj} of KATRIN data. We consider these results more accurate than the collaboration's result, since we do not allow for any parameters to go into an unphysical region of the parameter space in the marginalization process. The bounds become rapidly weak beyond $\Delta m^2_{41}\sim10^3~\textrm{eV}^2$ because the decay spectrum is only available up to 40~eV away from the end point. 

\section{Combination of short-baseline reactor data and Tritium constraints}
\label{sec:reactors-tritium}

Before the on-going KATRIN experiment,
there have been many experiments on the search of the effects of neutrino masses
in $\beta$ decays,
following the basic idea presented by
Enrico Fermi in 1933
in the same article in which he formulated the theory
of weak interactions~\cite{Fermi:1934sk}.
The strongest limits have been obtained in Tritium $\beta$-decay
experiments
(see, e.g., the recent review in Ref.~\cite{Formaggio:2021nfz}).
Some experiments have measured the electron spectrum
in a large energy interval below the end point.
Since none of these experiments found a deviation of the electron spectrum
from that predicted in the case of massless neutrino,
they produced bounds on the mixing of a heavy neutrino with the electron neutrino
which depend on the mass of the heavy neutrino.
In the framework of 3+1 active-sterile neutrino mixing,
the results of these experiments imply bounds on
$|U_{e4}|^2$ and the corresponding
$ \sin^2\!2\vartheta_{ee} = 4 |U_{e4}|^2 ( 1 - |U_{e4}|^2 ) $
which depend on the value of $m_{4}$.

In this section we present the results of the combined analysis of
short-baseline reactor data and the Tritium data.
We assume that the three standard neutrino masses
$m_{1}$,
$m_{2}$, and
$m_{3}$
are much lighter than $m_{4}$,
so that
$\Delta m_{41}^2 \simeq m_{4}^2$.

Besides the KATRIN bounds presented in Section~\ref{sec:KATRIN},
we consider the bounds of the previous Tritium experiments
Mainz~\cite{Kraus:2012he}
and
Troitsk~\cite{Belesev:2012hx,Belesev:2013cba}.
For these experiments,
we use the results obtained in Ref.~\cite{Giunti:2012bc}
and already used in several papers (e.g., Refs.~\cite{Gariazzo:2017fdh,Giunti:2019fcj}).
The reason for considering also the
Troitsk bound
is illustrated in Fig.~\ref{fig:rates-ratios-tritium},
where one can see that the Troitsk bound is competitive with the KATRIN limit
at large values of $\Delta m_{41}^2$ and cannot be neglected.
For completeness we consider also the less strong
Mainz bound,
which is shown in Fig.~\ref{fig:rates-ratios-tritium},
because Mainz and Troitsk have been contemporary and competitive experiments
of the generation before KATRIN.

In Fig.~\ref{fig:rates-ratios-tritium}
we show the combined $2\sigma$ and $3\sigma$ allowed regions
in the
($\sin^2\!2\vartheta_{ee},\Delta{m}^2_{41}$)
plane obtained from
the data of the
reactor rate experiments with different flux models,
the spectral ratio experiments,
and the three Tritium experiments
KATRIN, Troitsk and Mainz.
Comparing Fig.~\ref{fig:rates-ratios-tritium-neos17-3s}
with Fig.~\ref{fig:rates-ratios-neos17-3s},
one can see that when the NEOS/Daya Bay data are used
(RSRF(N/DB) fit)
the large-$\Delta m_{41}^2$ $3\sigma$ allowed regions in the cases of the
HM and HKSS reactor flux models
are limited by the bounds of the Tritium experiments.
When the NEOS/RENO data are used
(RSRF(N/R) fit),
all the reactor flux models give $3\sigma$ exclusion curves.
Comparing Figs.~\ref{fig:rates-ratios-tritium-neos21-2s}
and~\ref{fig:rates-ratios-tritium-neos21-3s}
with Figs.~\ref{fig:rates-ratios-neos21-2s}
and~\ref{fig:rates-ratios-neos21-3s},
one can see that the Tritium data allow us to improve
the exclusion of large mixing for
large values of $\Delta m_{41}^2$.

The addition of the Tritium bounds increases the tension between
the reactor data and the Gallium data in the framework
of short-baseline active-sterile oscillations.
This can be seen by comparing the values of the parameter goodness of fit
in Tabs.~\ref{tab:rea-gal-pgf}
and~\ref{tab:rea-tri-gal-pgf}.

For completeness,
in Fig.~\ref{fig:see-gal-tri-3s}
we show the regions in the
($\sin^2\!2\vartheta_{ee},\Delta{m}^2_{41}$)
plane which are allowed at $3\sigma$ by the
combined analysis
of the Gallium data with different cross sections
and the data of the Mainz, Troitsk, and KATRIN Tritium experiments.
Comparing the combined allowed regions with those in
Fig.~\ref{fig:rates-ratios-tritium},
one can see that the Tritium bounds
reduce the Gallium allowed regions at large values of $\Delta{m}^2_{41}$,
limiting the $3\sigma$ allowed regions at
$\Delta{m}^2_{41} \lesssim 30~\text{eV}^2$.
This figure is useful for discussions of the Gallium Anomaly
in which the bounds from other experiments
(i.e. those of the reactor experiments discussed above
and the solar neutrino bound discussed in Section~\ref{sec:solar})
are not considered.

%%%%%%%%%%%%%%%%%%%%%%%%%%%%%%%%%%%%%%%%%%%%%%%%%%%%%%%%

\begin{figure*}
\centering
\subfigure[]{ \label{fig:solar-bound-a}
\includegraphics[width=0.48\linewidth]{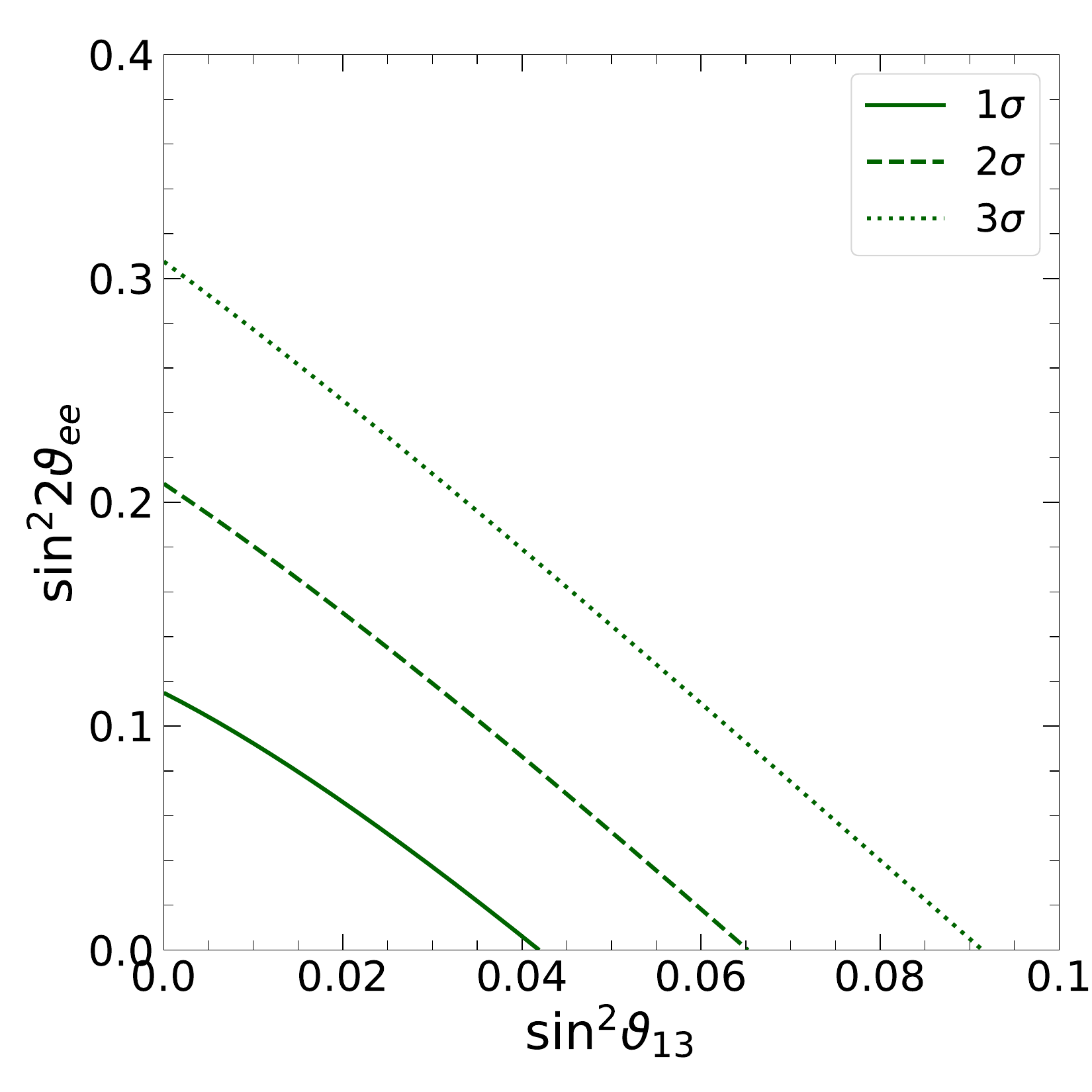}
}
\subfigure[]{ \label{fig:solar-bound-b}
\includegraphics[width=0.48\textwidth]{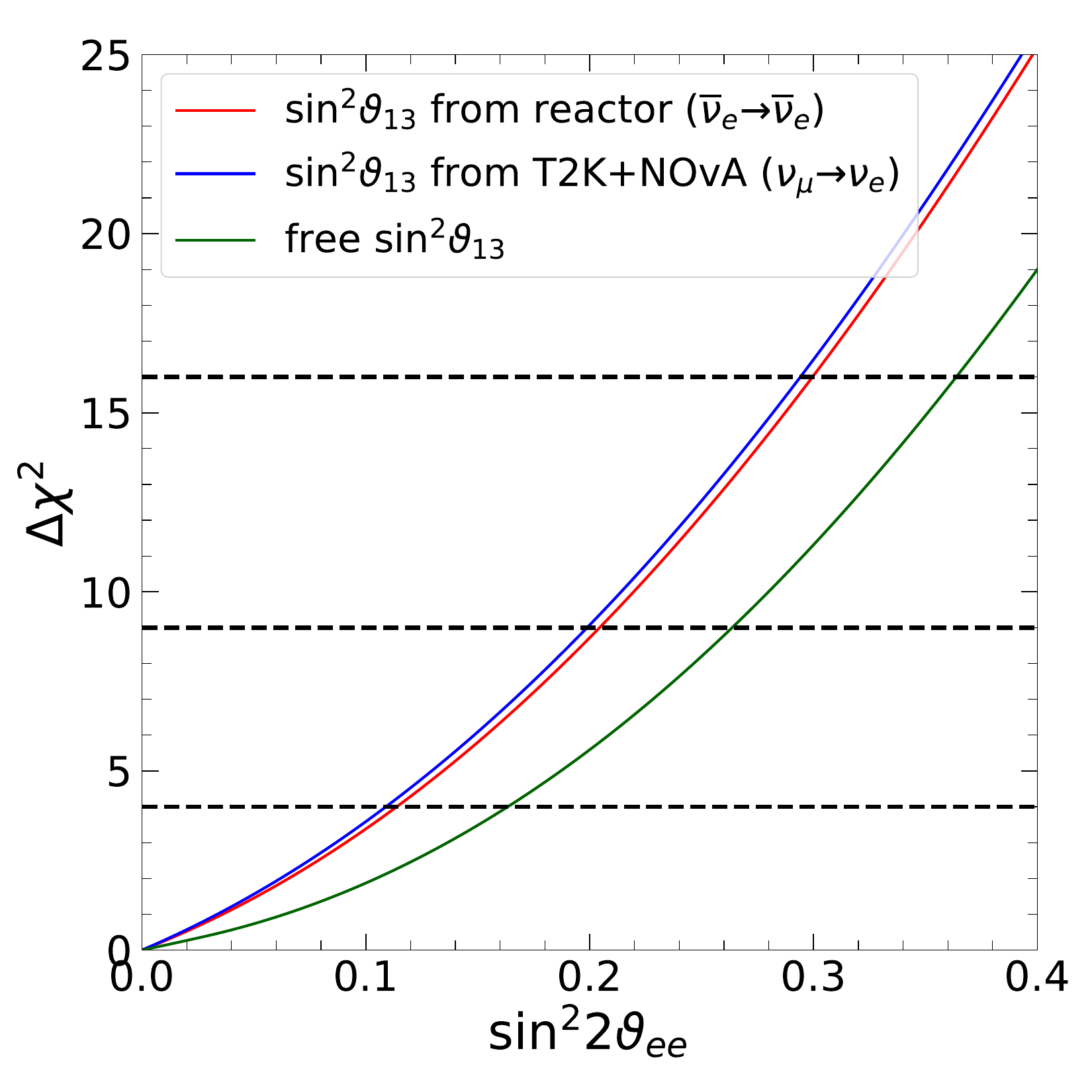}
}
\caption{\label{fig:solar-bound} 
\subref{fig:solar-bound-a} 
Allowed region in the ($\sin^2{\vartheta_{13}}$, $\sin^2{2\vartheta_{ee}}$) plane from the global solar data with free $\sin^2{\vartheta_{12}}$.
\subref{fig:solar-bound-b} The one-dimensional $\Delta \chi^2$ profiles of $\sin^2{2\vartheta_{ee}}$, where the red line corresponds to the analysis including the $\sin^2{\vartheta_{13}}$ constraint from PDG 2020~\cite{ParticleDataGroup:2020ssz}, 
the blue line is for the case of $\sin^2{\vartheta_{13}}$ constrained by neutrino data from T2K and NOvA~\cite{Tortola:2020ncu},
and the green line corresponds to the case with unconstrained $\sin^2{\vartheta_{13}}$. 
}
\end{figure*}

\begin{table*}
\centering
\begin{tabular}{|c|cc|cc|cc|}
\cline{2-7}
\multicolumn{1}{c|}{}
&
\multicolumn{2}{c|}{Solar-only}
&
\multicolumn{2}{c|}{S+$\vartheta_{13}$(T\&N)}
&
\multicolumn{2}{c|}{S+$\vartheta_{13}$(R)}
\\
\multicolumn{1}{c|}{}
&
$\Delta\chi^{2}_{\text{PG}}$
&
$\text{GoF}_{\text{PG}}$
&
$\Delta\chi^{2}_{\text{PG}}$
&
$\text{GoF}_{\text{PG}}$
&
$\Delta\chi^{2}_{\text{PG}}$
&
$\text{GoF}_{\text{PG}}$
\\
\hline
Ground State
&
$7.31$
&
$2.6\%$
&
$10.65$
&
$0.49\%$
&
$10.32$
&
$0.57\%$
\\
\hline
Bahcall
&
$10.30$
&
$0.58\%$
&
$14.14$
&
$0.085\%$
&
$13.78$
&
$0.1\%$
\\
\hline
Kostensalo
&
$9.03$
&
$1.1\%$
&
$12.79$
&
$0.17\%$
&
$12.43$
&
$0.2\%$
\\
\hline
Semenov
&
$12.70$
&
$0.17\%$
&
$17.24$
&
$0.018\%$
&
$16.83$
&
$0.022\%$
\\
\hline
\end{tabular}

\caption{ \label{tab:sun-gal-pgf}
$\chi^2$ difference
$\Delta\chi^{2}_{\text{PG}}$
of the parameter goodness of fit test~\cite{Maltoni:2003cu}
applied to the comparison of the solar neutrino bounds
and the neutrino oscillation explanation of the Gallium data
with the four Gallium detection cross sections
in Tab.~\ref{tab:gallium_cross_sections}.
The values of the corresponding parameter goodness of fit
$\text{GoF}_{\text{PG}}$
are calculated
with two degrees of freedom corresponding to the two
common oscillation parameters
$\sin^2\!2\vartheta_{ee}$ and $\Delta{m}^2_{41}$.
The Solar-only,
S+$\vartheta_{13}$(T\&N), and
S+$\vartheta_{13}$(R)
column titles refer, respectively, to
the free $\sin^2\vartheta_{13}$ analysis,
the analysis with $\sin^2\vartheta_{13}$ constrained by neutrino data from T2K and NOvA,
and the analysis with $\sin^2\vartheta_{13}$ constrained by reactor data.
}
\end{table*}

\section{The solar neutrino bound}
\label{sec:solar}

Solar neutrinos provide a robust method to constrain $\nu_e$ disappearance at short baselines~\cite{Giunti:2009xz,Giunti:2012tn,Palazzo:2011rj,Palazzo:2011rj,Palazzo:2012yf,Goldhagen:2021kxe,Dentler:2018sju}. 
The solar neutrino data include the measurements from three distinct channels, i.e., charged current (CC) interactions~\cite{Cleveland:1998nv,Kaether:2010ag,SAGE:2009eeu,SNO:2011hxd}, neutral current (NC) interactions~\cite{SNO:2011hxd}, and electron-scattering (ES) interactions~\cite{SNO:2011hxd,Super-Kamiokande:2005wtt,Super-Kamiokande:2008ecj,Super-Kamiokande:2010tar,Bellini:2011rx,Borexino:2008fkj,BOREXINO:2014pcl}. Each of the detection channels is sensitive to a different combination of the electron neutrino $\nu_{e}$,
and the other active neutrinos $\nu_{\mu}$,
$\nu_{\tau}$.
In the framework of the 3+1 neutrino mixing scheme, both the electron neutrino survival probability $P_{ee}$ and the electron-to-sterile neutrino transitional probability $P_{es}$ can be constrained with these CC, NC and ES measurements.
In this work, we employ the simplified treatment of the solar neutrino analysis proposed in Ref.~\cite{Goldhagen:2021kxe}, where only the asymptotic values of $P_{ee}$ and $P_{es}$ at the high and low energies are employed to constrain the relevant mixing parameters. This method has been proven to consistently reproduce the full data analysis treatment~\cite{Goldhagen:2021kxe}, since the eV-scale sterile neutrinos do not
modify the energy dependence of the MSW resonance region in the oscillation probabilities~\cite{Giunti:2009xz}.

In the 3+1 mixing scheme, the solar neutrino oscillation probabilities $P_{ee}$ and $P_{es}$ can be written as
\begin{align}
  P_{ee} =  \sum_{k=1}^4 |U_{ek}^m|^2|U_{ek}|^2 \,,\quad %\label{eq.Pee}\\
  P_{es} =   
    \sum_{k=1}^4 |U_{ek}^m|^2|U_{sk}|^2 \,, \label{eq.Pex}
\end{align}
where $U_{ek}^m$ is the mixing element in matter at the production point of solar neutrinos. By using the parametrization of the mixing matrix $U$ in Ref.~\cite{Giunti:2009xz}, 
the low energy (LE) and high energy (HE) asymptotic values of $P_{ee}$ and $P_{es}$ can be easily calculated as 
\begin{align}
    P_{ee}^{\rm{LE}} & = c_{12}^4 c_{13}^4 c_{14}^4 + s_{12}^4 s_{13}^4 c_{14}^4 + s_{13}^4 c_{14}^4 + s_{14}^4 \label{eq:PeeLE} \, , \\
    P_{ee}^{\rm{HE}} & = s_{12}^2 c_{13}^4 c_{14}^4 +  s_{13}^4 c_{14}^4 + s_{14}^4 \label{eq:PeeHE} \, , \\
    % P_{es}^{\rm{LE}} & =  c_{12}^4 c_{13}^4 s_{14}^2 c_{14}^2 + s_{12}^4 c_{13}^4 s_{14}^2 c_{14}^2 + s_{13}^4 s_{14}^2 c_{14}^2 + s_{14}^2 c_{14}^2 \, , \\
    % P_{es}^{\rm{HE}} & =  s_{12}^2 c_{13}^4 s_{14}^2 c_{14}^2 + s_{13}^4 s_{14}^2 c_{14}^2 +s_{14}^2c_{14}^2 \, ,
    P_{es}^{\rm{LE}} & =  \left( c_{12}^4 c_{13}^4  + s_{12}^4 c_{13}^4 + s_{13}^4 + 1 \right) s_{14}^2 c_{14}^2 \label{eq:PesLE}\, , \\
    P_{es}^{\rm{HE}} & = \left( s_{12}^2 c_{13}^4  + s_{13}^4 + 1 \right)s_{14}^2 c_{14}^2 \label{eq:PesHE} \, ,
\end{align}
where $s_{ij}\equiv\sin\vartheta_{ij}$ and $c_{ij}\equiv\cos\vartheta_{ij}$ ($1\leq i,j\leq4$) have been used, and $\vartheta_{12}^m({\rm LE})=\vartheta_{12}$ and $\vartheta_{12}^m({\rm HE}) = \pi/2$ are employed. Before doing the numerical analysis, we make the following comments on the analysis method:
\begin{description}
\item[A]
$\vartheta_{24}=0$ and $\vartheta_{34}=0$ are used in the above calculations, which have been proven to have no visible effects~\cite{Goldhagen:2021kxe,Palazzo:2011rj,Palazzo:2012yf,Dentler:2018sju} on the solar neutrino data when considering the corresponding constraints of $\vartheta_{24}$ and $\vartheta_{34}$ from measurements of accelerator $\nu_{\mu}$ disappearance~\cite{Dydak:1983zq,SciBooNE:2011qyf}, atmospheric neutrinos~\cite{Super-Kamiokande:2014ndf,IceCube:2020phf} and long-baseline accelerator NC interactions~\cite{MINOS:2011ysd,NOvA:2017geg}.

\item[B]
KamLAND data~\cite{KamLAND:2013rgu} is not included in the analysis. But the value of $\Delta m^2_{21}$ has been implicitly assumed to not significantly deviate from the current allowed region of the large mixing angle solution of the MSW resonance, and thus does not alter the asymptotic values of oscillation probabilities. This assumption is seen to be valid by considering the allowed range of $\Delta m^2_{21}$ from the solar-only analysis~\cite{deSalas:2020pgw,Esteban:2020cvm,Capozzi:2021fjo,Vissani:2017wvk}.
\end{description}

The $\chi^2$ function employed in this analysis is
\begin{align}
    \chi^2 \left(s_{12}^2, s_{13}^2, \sin^2\!2\vartheta_{ee}\right) = \sum_{a,b} (O_a - P_a) V_{ab}^{-1} (O_b - P_b) \, ,
\end{align}
where $P_a=\left(P_{ee}^{\rm{LE}}, P_{ee}^{\rm{HE}}, P_{e\mu}^{\rm{LE}}+P_{e\tau}^{\rm{LE}}, P_{e\mu}^{\rm{HE}}+P_{e\tau}^{\rm{HE}}\right)$, $O_a$ are the measurements of the asymptotic values of oscillation probabilities from
the full solar neutrino analysis, and $V_{ab}$ is the covariance matrix taking into account possible correlations~\cite{Goldhagen:2021kxe}.
The transition probability
$P_{e\mu}+P_{e\tau}$ is given by
$1 - P_{ee} - P_{es}$.
In the following, the notation of $\sin^2\!2\vartheta_{ee} \equiv 4 s_{14}^2(1-s_{14}^2)$ will be used for consistency with previous sections.

Figure~\ref{fig:solar-bound-a} illustrates the allowed regions in the $(\sin^2\vartheta_{13}, \sin^22\vartheta_{ee})$ plane obtained from the analysis of solar neutrino data where $\sin^2\vartheta_{12}$ has been marginalized over. The green solid, dashed and dotted lines show the allowed regions at $1\sigma$, $2\sigma$ and $3\sigma$ confidence levels, respectively. One can see that $\sin^2\!2\vartheta_{ee}$ and $\sin^2\vartheta_{13}$ are essentially degenerate as already revealed in Eqs.~(\ref{eq:PeeLE}-\ref{eq:PesHE}). The allowed region is located in the lower-left corner of Fig.~\ref{fig:solar-bound-a}, which indicates that $\sin^2\!2\vartheta_{ee}\gtrsim 0.2$ is practically ruled out at the $2\sigma$ confidence level, independently of the value of $\sin^2\vartheta_{13}$. The bound on $\sin^2\!2\vartheta_{ee}$ becomes more stringent as $\sin^2\vartheta_{13}$ increases.

In Fig.~\ref{fig:solar-bound-b} the one-dimensional $\chi^2$ profiles of $\sin^2\!2\vartheta_{ee}$ obtained from solar neutrino data are illustrated. The green line is obtained using only solar neutrino data, in which no external information on $\vartheta_{13}$ has been imposed. As can be seen, this bound is already in tension with the preferred region of the Gallium experiments discussed in Section~\ref{sec:gallium}. In the first column of Tab.~\ref{tab:sun-gal-pgf} we report the parameter goodness of fit taking into account only solar neutrino data. As can be seen the $\text{GoF}_{\text{PG}}$ is below 1\% for the Bahcall and Semenov cross section models. Using the Kostensalo cross section model, we find $\text{GoF}_{\text{PG}} \sim 1\%$. Only the extreme case of the Ground State cross section gives $\text{GoF}_{\text{PG}}>1\%$.

The red line in Fig.~\ref{fig:solar-bound-b} represents the result when taking into account the reactor antineutrino constraint on $\sin \vartheta_{13}$ from PDG 2020~\cite{ParticleDataGroup:2020ssz}, i.e. $\sin^2 \vartheta_{13} = (2.20 \pm 0.07)\times 10^{-2}$. We find that values $\sin^2\!2\vartheta_{ee} > 0.1$ are excluded at 2$\sigma$ and therefore the tension with Gallium data is further increased. As shown in the right column of Tab.~\ref{tab:sun-gal-pgf}, the tension becomes unacceptable for all of the cross section models.

The tension between reactor antineutrino data and Gallium neutrino data was already discussed in Ref.~\cite{Giunti:2010zs}. Therein the authors proposed a CPT-violating solution, where the parameters characterizing neutrino oscillations could differ from their antineutrino counterparts~\cite{Barenboim:2002tz}. We show that using the newest solar neutrino data, this solution is not viable anymore. The blue line in Fig.~\ref{fig:solar-bound-b} shows the result of our solar neutrino analysis by considering the bound on $\vartheta_{13}$ obtained from the analysis of the neutrino mode data collected in T2K and NOvA~\cite{Tortola:2020ncu}, i.e., $\sin^2 \vartheta_{13} = (2.60^{+ 1.03}_{-0.48})\times 10^{-2}$. One can see that the bound is even slightly stronger than the former one. This is due to the fact that T2K+NOvA in neutrino mode measure a best fit value which is slightly larger than the reactor antineutrino measurement and, as can be seen from Fig.~\ref{fig:solar-bound-a}, larger values of $\sin^2\vartheta_{13}$ require smaller values of $\sin^22\vartheta_{ee}$. As a result the tension with Gallium data is basically the same as in the case where the reactor constraint on $\vartheta_{13}$ is used, which can also be seen by comparing the central and right columns of Tab.~\ref{tab:sun-gal-pgf}. As a consequence, the CPT-violating solution~\cite{Giunti:2010zs} is ruled out by current solar data in combination with the constraint on $\vartheta_{13}$ from the neutrino data of T2K and NOvA~\cite{Tortola:2020ncu}.

%%%%%%%%%%%%%%%%%%%%%%%%%%%%%%%%%%%%%%%%%%%%%%%%%%%%%%%%

\begin{table}
\centering
\begin{tabular}{|c|c|c|c|c|}
\cline{2-5}
\multicolumn{1}{c|}{}
&
\multicolumn{4}{c|}{Global RSRF(N/DB) Fit}
\\
\cline{2-5}
\multicolumn{1}{c|}{}
&
HM
&
HKSS
&
EF
&
KI
\\
\hline
$\chi^2_{\text{min}}$
&
$393.5$
&
$395.2$
&
$391.2$
&
$391.4$
\\
GoF
&
$43\%$
&
$40\%$
&
$46\%$
&
$46\%$
\\
$(\sin^2 2\vartheta_{ee})_{\text{b.f.}}$
&
$0.022$
&
$0.022$
&
$0.022$
&
$0.022$
\\
$(\Delta m_{41}^2)_{\text{b.f.}} / \text{eV}^2$
&
$1.29$
&
$1.29$
&
$1.29$
&
$1.29$
\\
$\Delta\chi^2_{4\nu\mbox{-}3\nu}$
&
$13.8$
&
$14.1$
&
$12.6$
&
$12.9$
\\
$n\sigma_{4\nu\mbox{-}3\nu}$
&
$3.3$
&
$3.3$
&
$3.1$
&
$3.2$
\\
\hline
\multicolumn{1}{c|}{}
&
\multicolumn{4}{c|}{Global RSRF(N/R) Fit}
\\
\cline{2-5}
\multicolumn{1}{c|}{}
&
HM
&
HKSS
&
EF
&
KI
\\
\hline
$\chi^2_{\text{min}}$
&
$386.5$
&
$388.3$
&
$384.0$
&
$384.2$
\\
GoF
&
$53\%$
&
$50\%$
&
$56\%$
&
$56\%$
\\
$(\sin^2 2\vartheta_{ee})_{\text{b.f.}}$
&
$0.017$
&
$0.019$
&
$0.017$
&
$0.017$
\\
$(\Delta m_{41}^2)_{\text{b.f.}} / \text{eV}^2$
&
$1.32$
&
$1.32$
&
$1.32$
&
$1.32$
\\
$\Delta\chi^2_{4\nu\mbox{-}3\nu}$
&
$10.1$
&
$10.3$
&
$9.1$
&
$9.3$
\\
$n\sigma_{4\nu\mbox{-}3\nu}$
&
$2.7$
&
$2.8$
&
$2.6$
&
$2.6$
\\
\hline
\end{tabular}

\caption{ \label{tab:rea-tri-sun-fit}
Results of the global $\nu_e$ and $\bar\nu_e$ disappearance fits
obtained using
NEOS/Daya Bay
(RSRF(N/DB))
or
NEOS/RENO
(RSRF(N/R))
data.
The column titles HM, HKSS, EF, and KI
refer to the four reactor neutrino fluxes discussed in Section~\ref{sec:rates}.
The table show
the minimum value $\chi^2_{\text{min}}$ of $\chi^2$,
the goodness of fit (GoF) with 390 degrees of freedom
(the number of data points minus two,
corresponding to the two fitted mixing parameters),
the best-fit values of
$\sin^2\!2\vartheta_{ee}$
and
$\Delta m_{41}^2$,
the $\chi^2$ difference $\Delta\chi^2_{4\nu\mbox{-}3\nu}$
between the 3+1 $4\nu$ fit and the $3\nu$ fit, and
the statistical significance
($n\sigma_{4\nu\mbox{-}3\nu}$)
of the corresponding indication in favor of 3+1 $4\nu$ mixing.
}
\end{table}

\begin{figure*}
\centering
\subfigure[]{ \label{fig:rates-ratios-tritium-sun-neos17-2s}
\includegraphics[width=0.48\linewidth]{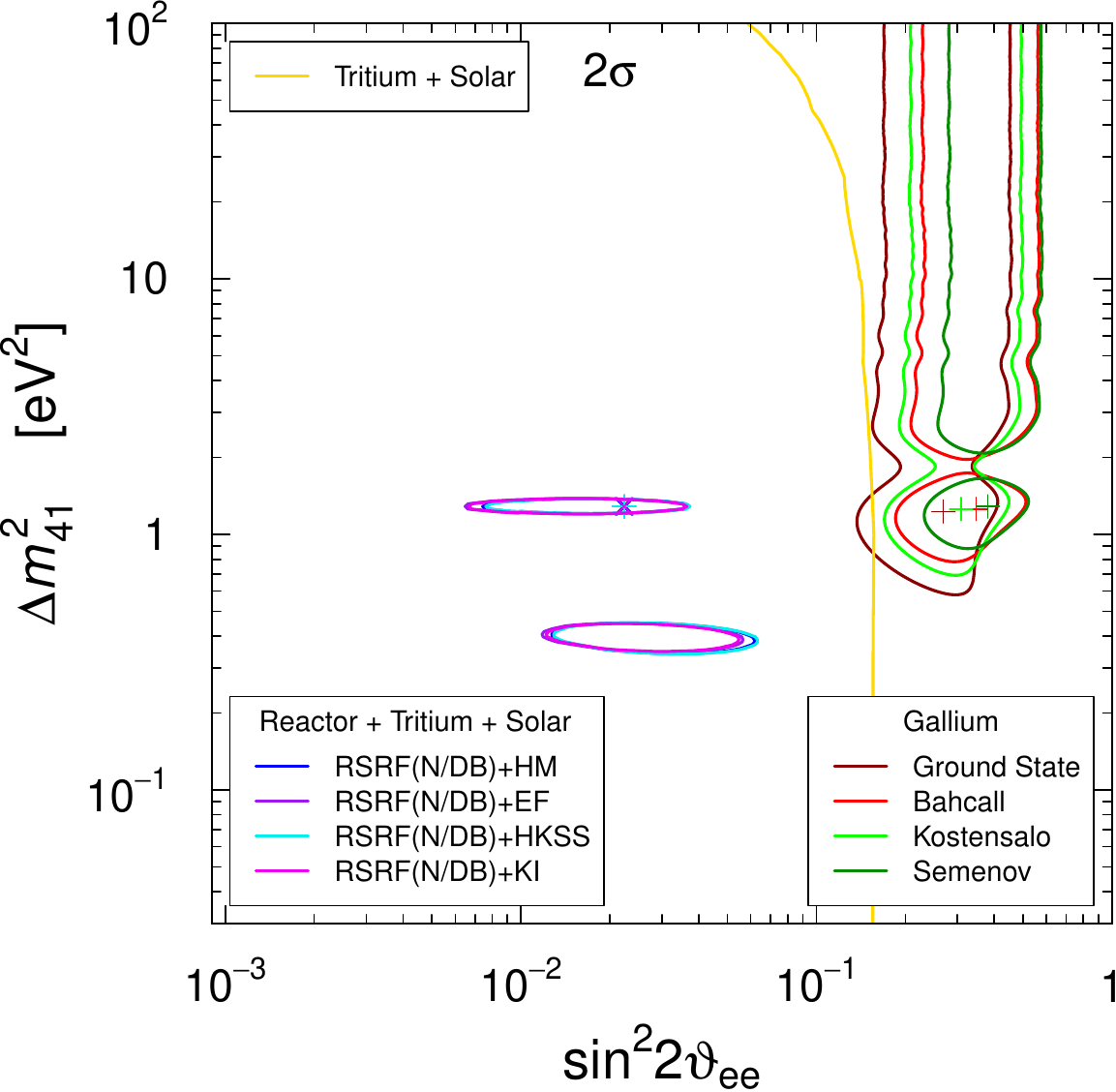}
}
\subfigure[]{ \label{fig:rates-ratios-tritium-sun-neos21-2s}
\includegraphics[width=0.48\textwidth]{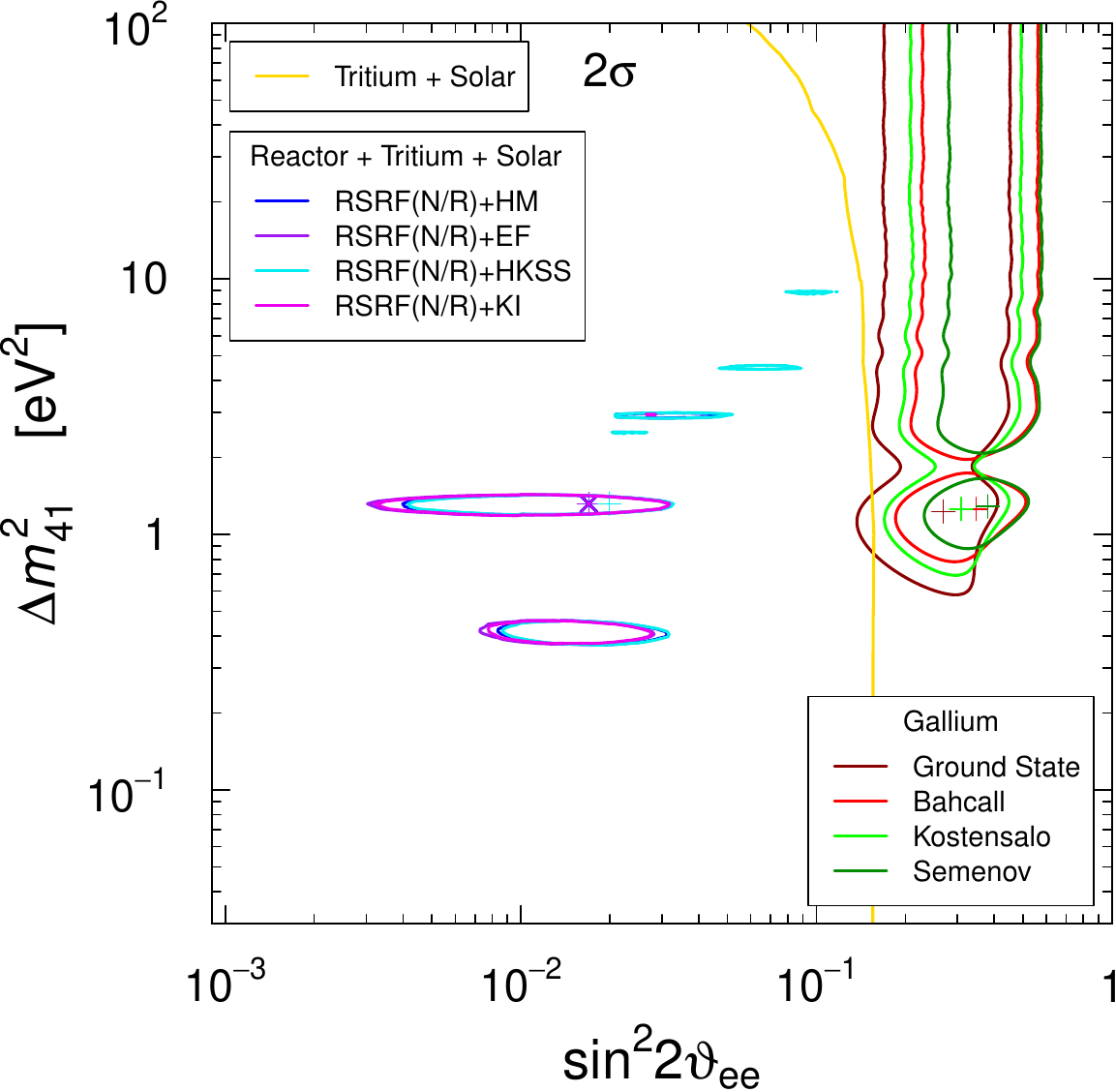}
}
\\
\subfigure[]{ \label{fig:rates-ratios-tritium-sun-neos17-3s}
\includegraphics[width=0.48\linewidth]{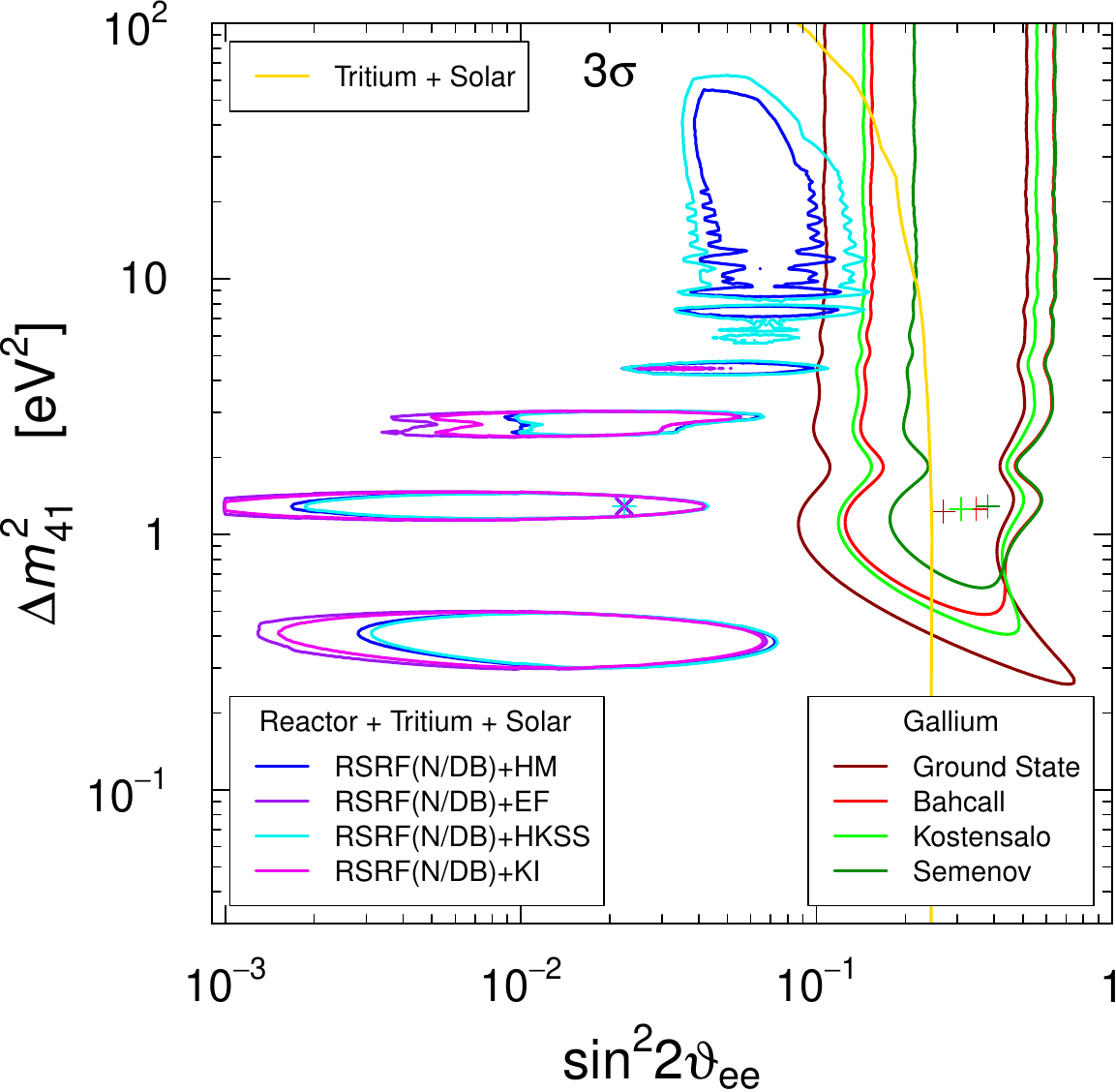}
}
\subfigure[]{ \label{fig:rates-ratios-tritium-sun-neos21-3s}
\includegraphics[width=0.48\textwidth]{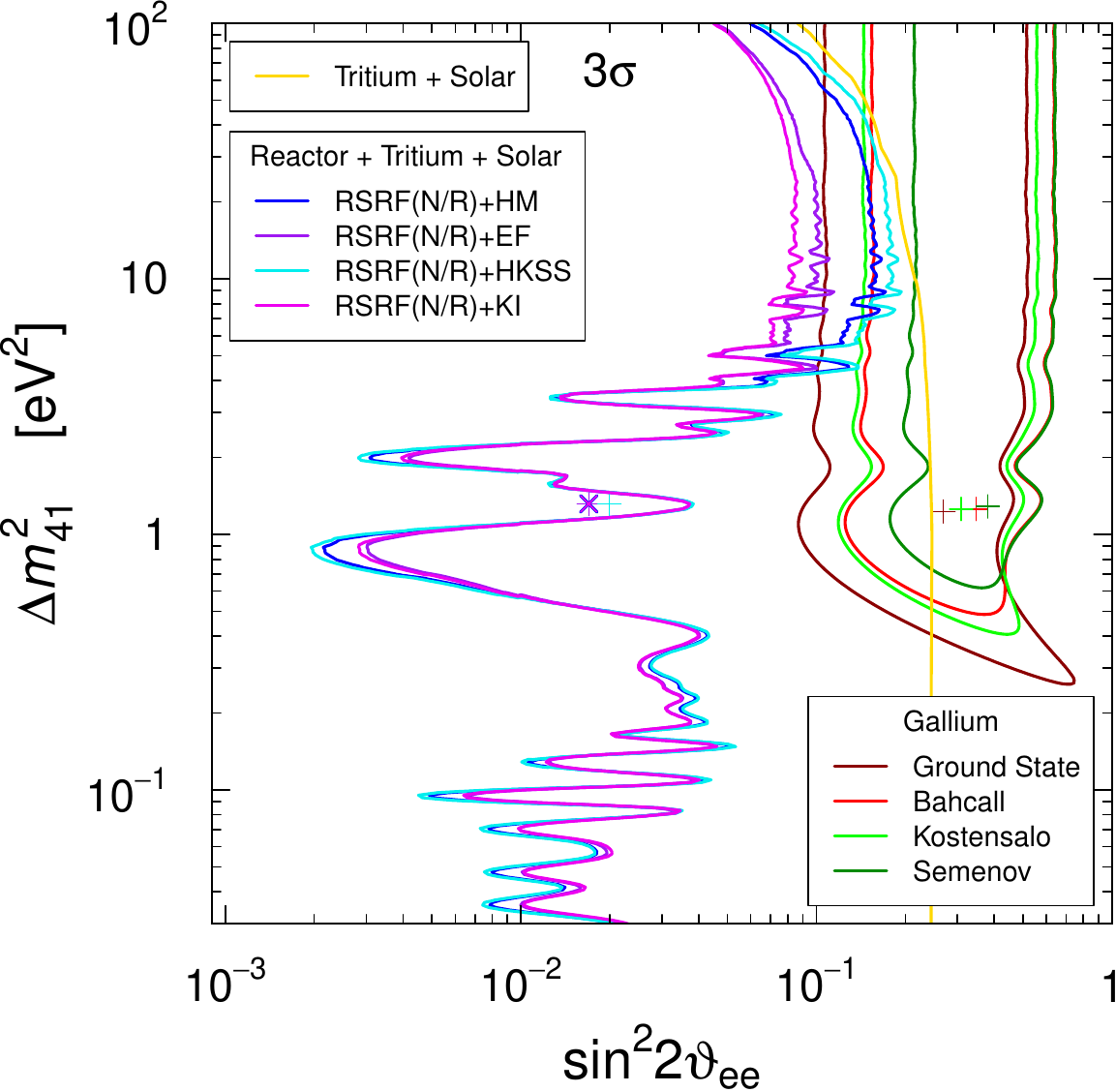}
}
\caption{\label{fig:rates-ratios-tritium-sun}
Comparison of the contours delimiting the
[\subref{fig:rates-ratios-tritium-sun-neos17-2s} and \subref{fig:rates-ratios-tritium-sun-neos21-2s}]
$2\sigma$ and
[\subref{fig:rates-ratios-tritium-sun-neos17-3s} and \subref{fig:rates-ratios-tritium-sun-neos21-3s}]
$3\sigma$
allowed regions in the
($\sin^2\!2\vartheta_{ee},\Delta{m}^2_{41}$)
plane obtained from the combined analysis of the data of the
reactor rate experiments with different flux models,
the spectral ratio experiments,reactor 
the Tritium experiments,
and the solar bound
with those obtained from the  Gallium data with different cross sections.
Also shown is the $3\sigma$ bound obtained from the
combination of the Tritium and solar bounds.
The figures differ by the use of
[\subref{fig:rates-ratios-tritium-sun-neos17-2s} and \subref{fig:rates-ratios-tritium-sun-neos17-3s}]
NEOS/Daya Bay~\cite{NEOS:2016wee}
or
[\subref{fig:rates-ratios-tritium-sun-neos21-2s} and \subref{fig:rates-ratios-tritium-sun-neos21-3s}]
NEOS/RENO~\cite{RENO:2020hva}
spectral ratio data.
The best-fit points are indicated by crosses.
}
\end{figure*}

\section{Global $\nu_e$ and $\bar\nu_e$ disappearance analysis}
\label{sec:global}

In this Section we present the results of the global analysis of the
$\nu_e$ and $\bar\nu_e$ disappearance data
in the framework of 3+1 active-sterile neutrino mixing.
The data that we consider are
the reactor rates
discussed in Section~\ref{sec:rates},
the reactor spectral ratio data
discussed in Section~\ref{sec:ratios},
the Tritium limits
discussed in Sections~\ref{sec:KATRIN} and~\ref{sec:reactors-tritium},
and the solar neutrino bound
discussed in Section~\ref{sec:solar}.
We will discuss the global tension between the data that we consider
and the Gallium data if the Gallium Anomaly
is considered as due to 3+1 active-sterile neutrino mixing.
Since the tension is very strong,
we cannot include the Gallium data in a global fit
in the framework of 3+1 active-sterile neutrino mixing,
and we presume that the Gallium Anomaly is due to other reasons.

The results of the global fits
that we obtained with different data sets
(NEOS/Daya Bay or NEOS/RENO)
and different reactor flux models
are listed in Tab.~\ref{tab:rea-tri-sun-fit}.
One can see that the goodness of fit is high.
There is a 3.1--3.3$\sigma$
indication in favor of 3+1 active-sterile neutrino mixing
in the global fits with the NEOS/Daya Bay data.
The indication decreases to 2.6--2.8$\sigma$
if the NEOS/RENO are used.
The values of the best-fit points are in any case around
$\sin^2\!2\vartheta_{ee} \simeq 0.02$
and
$\Delta m_{41}^2 \simeq 1.3~\text{eV}^2$.

Figure~\ref{fig:rates-ratios-tritium-sun}
shows the $2\sigma$ and $3\sigma$ allowed regions in the
($\sin^2\!2\vartheta_{ee},\Delta{m}^2_{41}$) plane
obtained from the global fits
with different neutrino flux models
and considering either the
NEOS/Daya Bay~\cite{NEOS:2016wee}
spectral ratio
(Figs.~\ref{fig:rates-ratios-tritium-sun-neos17-2s}
and~\ref{fig:rates-ratios-tritium-sun-neos17-3s})
or the
NEOS/RENO~\cite{RENO:2020hva}
spectral ratio
(Figs.~\ref{fig:rates-ratios-tritium-sun-neos21-2s}
and~\ref{fig:rates-ratios-tritium-sun-neos21-3s}).
Comparing Fig.~\ref{fig:rates-ratios-tritium-sun}
with Fig.~\ref{fig:rates-ratios-tritium},
one can see that the addition of the solar bound
has the effect of reducing the allowed regions for
$\sin^2\!2\vartheta_{ee} \gtrsim 0.1$.
In the case of the fit with the NEOS/Daya Bay spectral ratio,
comparing Figs.~\ref{fig:rates-ratios-tritium-sun-neos17-2s}
and~\ref{fig:rates-ratios-tritium-sun-neos17-3s}
with Figs.~\ref{fig:rates-ratios-tritium-neos17-2s}
and~\ref{fig:rates-ratios-tritium-neos17-3s},
one can see that the solar bound has the effect of moving the best-fit points
obtained with the HM and HKSS reactor neutrino fluxes
from the allowed island at
$\Delta m_{41}^2 \simeq 0.4 \, \text{eV}^2$
to the
allowed island at
$\Delta m_{41}^2 \simeq 1.3 \, \text{eV}^2$,
leading to an approximate coincidence of the best-fit points
obtained with all the four reactor neutrino fluxes.
From Figs.~\ref{fig:rates-ratios-tritium-sun-neos21-2s}
and~\ref{fig:rates-ratios-tritium-sun-neos21-3s}
one can see that in the analysis with the
NEOS/RENO spectral ratio
there is a similar approximate coincidence of the best-fit points
obtained with all the four reactor neutrino fluxes.

\begin{table*}
\centering
\begin{tabular}{|c|cc|cc|cc|cc|}
\cline{2-9}
\multicolumn{1}{c|}{}
&
\multicolumn{8}{c|}{Global Fit: RSRF(N/DB) + Reactor Rates + Tritium + Solar}
\\
\cline{2-9}
\multicolumn{1}{c|}{}
&
\multicolumn{2}{c|}{HM}
&
\multicolumn{2}{c|}{HKSS}
&
\multicolumn{2}{c|}{EF}
&
\multicolumn{2}{c|}{KI}
\\
\multicolumn{1}{c|}{}
&
$\Delta\chi^{2}_{\text{PG}}$
&
$\text{GoF}_{\text{PG}}$
&
$\Delta\chi^{2}_{\text{PG}}$
&
$\text{GoF}_{\text{PG}}$
&
$\Delta\chi^{2}_{\text{PG}}$
&
$\text{GoF}_{\text{PG}}$
&
$\Delta\chi^{2}_{\text{PG}}$
&
$\text{GoF}_{\text{PG}}$
\\
\hline
Ground State
&
$21.54$
&
$0.0021\%$
&
$19.51$
&
$0.0058\%$
&
$21.92$
&
$0.0017\%$
&
$21.90$
&
$0.0018\%$
\\
\hline
Bahcall
&
$25.99$
&
$0.00023\%$
&
$23.88$
&
$0.00065\%$
&
$26.13$
&
$0.00021\%$
&
$26.11$
&
$0.00021\%$
\\
\hline
Kostensalo
&
$25.05$
&
$0.00036\%$
&
$22.77$
&
$0.0011\%$
&
$27.62$
&
$0.0001\%$
&
$27.60$
&
$0.0001\%$
\\
\hline
Semenov
&
$32.52$
&
$0.0000087\%$
&
$29.93$
&
$0.000032\%$
&
$37.69$
&
$0.00000065\%$
&
$38.81$
&
$0.00000037\%$
\\
\hline
\multicolumn{1}{c|}{}
&
\multicolumn{8}{c|}{Global Fit: RSRF(N/R) + Reactor Rates + Tritium + Solar}
\\
\cline{2-9}
\multicolumn{1}{c|}{}
&
\multicolumn{2}{c|}{HM}
&
\multicolumn{2}{c|}{HKSS}
&
\multicolumn{2}{c|}{EF}
&
\multicolumn{2}{c|}{KI}
\\
\multicolumn{1}{c|}{}
&
$\Delta\chi^{2}_{\text{PG}}$
&
$\text{GoF}_{\text{PG}}$
&
$\Delta\chi^{2}_{\text{PG}}$
&
$\text{GoF}_{\text{PG}}$
&
$\Delta\chi^{2}_{\text{PG}}$
&
$\text{GoF}_{\text{PG}}$
&
$\Delta\chi^{2}_{\text{PG}}$
&
$\text{GoF}_{\text{PG}}$
\\
\hline
Ground State
&
$17.61$
&
$0.015\%$
&
$15.53$
&
$0.042\%$
&
$22.56$
&
$0.0013\%$
&
$22.66$
&
$0.0012\%$
\\
\hline
Bahcall
&
$22.07$
&
$0.0016\%$
&
$19.90$
&
$0.0048\%$
&
$26.82$
&
$0.00015\%$
&
$26.80$
&
$0.00015\%$
\\
\hline
Kostensalo
&
$21.11$
&
$0.0026\%$
&
$18.77$
&
$0.0084\%$
&
$26.27$
&
$0.0002\%$
&
$28.45$
&
$0.000066\%$
\\
\hline
Semenov
&
$28.57$
&
$0.000062\%$
&
$25.93$
&
$0.00023\%$
&
$34.00$
&
$0.0000041\%$
&
$38.24$
&
$0.0000005\%$
\\
\hline
\end{tabular}

\caption{ \label{tab:rea-tri-sun-gal-pgf}
$\chi^2$ difference
$\Delta\chi^{2}_{\text{PG}}$
of the parameter goodness of fit test~\cite{Maltoni:2003cu}
applied to the comparison of the neutrino oscillation fits of the reactor rates, spectral ratio data, tritium data, and solar bound
with the Gallium data
using the four Gallium detection cross sections
in Tab.~\ref{tab:gallium_cross_sections}.
The values of the corresponding parameter goodness of fit
$\text{GoF}_{\text{PG}}$
are calculated
with two degrees of freedom corresponding to the two
common oscillation parameters
$\sin^2\!2\vartheta_{ee}$ and $\Delta{m}^2_{41}$.
The column titles RSRF(N/DB) and RSRF(N/R)
indicate, respectively,
the combined fits of NEOS/Daya Bay and NEOS/RENO data with those
of the other reactor spectral ratio experiments discussed in the text
(DANSS, PROSPECT, STEREO, and Bugey-3).
The column titles HM, HKSS, EF, and KI
refer to the four reactor neutrino fluxes discussed in Section~\ref{sec:rates}.
}
\end{table*}

The restrictions of the solar neutrino constraint
increase the tension with the results of the Gallium experiments
in the framework of 3+1 active-sterile neutrino mixing,
as one can see comparing the parameter goodness of fit
in Tab.~\ref{tab:rea-tri-sun-gal-pgf}
with those in Tab.~\ref{tab:rea-tri-gal-pgf},
which were obtained without the solar bound.
From Tab.~\ref{tab:rea-tri-sun-gal-pgf},
one can see that the global fit gives values of parameter goodness of fit
that are well below 1\% for all the fits with different data combinations.
The values of the parameter goodness of fit in
Tab.~\ref{tab:rea-tri-sun-gal-pgf}
quantify the tension between the
$3\sigma$ Gallium allowed regions and the $3\sigma$ global allowed regions,
which have only marginal overlaps
in Fig.~\ref{fig:rates-ratios-tritium-sun}.

In Fig.~\ref{fig:rates-ratios-tritium-sun} one can notice the curious
approximate coincidence of the values of $\Delta m_{41}^2$
for the global best-fit points
($(\Delta m_{41}^2)_{\text{b.f.}} = 1.3 \, \text{eV}^2$)
and the Gallium best-fit points
($(\Delta m_{41}^2)_{\text{b.f.}} = 1.3 \, \text{eV}^2$
for Bahcall, Kostensalo and Semenov, and
$(\Delta m_{41}^2)_{\text{b.f.}} = 1.2 \, \text{eV}^2$
for Ground State).
This coincidence has no meaning in the framework of
3+1 active-sterile neutrino mixing,
because the best-fit values of the unique mixing parameter
$\sin^2\!2\vartheta_{ee}$
are too different and the $3\sigma$ allowed regions
are well separated for
$\Delta m_{41}^2 \approx 1.3 \, \text{eV}^2$.

%%%%%%%%%%%%%%%%%%%%%%%%%%%%%%%%%%%%%%%%%%%%%%%%%%%%%%%%

\section{Summary and conclusions}
\label{sec:conclusions}

In this paper we discussed in a systematic way
the results of $\nu_{e}$ and $\bar\nu_{e}$ disappearance experiments
which are relevant for the hypothetical existence of short-baseline neutrino oscillations
due to 3+1 active-sterile neutrino mixing.
We started in Section~\ref{sec:gallium} with the analysis of the results of the Gallium source experiments,
motivated by the recent results of the BEST experiment~\cite{Barinov:2021asz,Barinov:2022wfh},
which revived the Gallium Anomaly confirming the results of the
GALLEX~\cite{GALLEX:1994rym,GALLEX:1997lja,Kaether:2010ag}
and
SAGE~\cite{Abdurashitov:1996dp,SAGE:1998fvr,Abdurashitov:2005tb,SAGE:2009eeu}
experiments.
We have shown that the explanation of the
Gallium Anomaly in the framework of 3+1 active-sterile neutrino mixing
requires rather large values of the mixing between $\nu_{e}$ and the new massive neutrino $\nu_{4}$
for all the cross section models of neutrino detection in the Gallium source experiments.
This means that $\nu_{4}$ is not almost entirely sterile,
as it would be required for considering 3+1 active-sterile mixing as a perturbation of standard three-neutrino mixing.
This is required for the explanation of the neutrino oscillations
observed in solar, atmospheric and long-baseline neutrino oscillation experiments.
In Section~\ref{sec:solar},
we have shown that the neutrino oscillation explanation of the Gallium Anomaly is in strong tension
with the solar bound on active-sterile neutrino mixing.

We also considered the results of
reactor neutrino experiments and we presented the results of 3+1 fits of the
measured rates and the ratios of spectra measured at different distances.
In Section~\ref{sec:rates},
we have shown that the measured reactor rates imply upper bounds for active-sterile mixing
that are in tension with the results of the Gallium experiments
for all the cross section models of neutrino detection in the Gallium source experiments
and all the reactor neutrino flux models.

In Section~\ref{sec:ratios},
we presented the results of a global fit of the most recent available data of the
reactor neutrino experiments which measured the ratio of the neutrino spectra
at different distances in order to probe neutrino oscillations
independently from the absolute values of the neutrino fluxes. Our analysis updates the one of Ref.~\cite{Berryman:2021yan} by taking into account the newest DANSS data which have important effects on the combined analysis of spectral ratio data, as discussed in Section~\ref{sec:ratios}.
We have shown that the results depend significantly
on the choice to consider either the
NEOS/Daya Bay~\cite{NEOS:2016wee}
or
NEOS/RENO~\cite{RENO:2020hva}
spectral ratio data.
In the case of the NEOS/Daya Bay,
there is a $3.1\sigma$ indication in favor of
short-baseline oscillations with best-fit parameter values
$\sin^2\!2\vartheta_{ee}=0.022$
and
$\Delta m_{41}^2 = 1.29 \, \text{eV}^2$,
which is driven by the overlap of the surrounding allowed regions of
NEOS/Daya Bay
and
DANSS~\cite{DANSS-ICHEP2022}.
On the other hand,
the overlap of the
NEOS/RENO
and
DANSS
allowed regions is smaller and leads to an indication in favor of
short-baseline oscillations of only $2.6\sigma$
with best-fit parameters values
$\sin^2\!2\vartheta_{ee}=0.017$
and
$\Delta m_{41}^2 = 1.32 \, \text{eV}^2$.
Although the NEOS/RENO comparison may be favored by the smaller systematic uncertainties which were estimated by the
NEOS and RENO collaborations using the fact that the two experiments
detect neutrino fluxes from similar reactors in the same complex,
we think that we cannot dismiss the NEOS/Daya Bay data,
which should be considered as a different measurement based on
the Daya Bay neutrino flux measurement.
Since we cannot combine the two measurements because the NEOS data would be double counted,
we remain with the ambiguity of the two different results
which hopefully will be solved by future measurements.

We have also shown that the results of the reactor spectral ratio experiments
are in tension with the neutrino oscillation explanation of the Gallium Anomaly
and the tension is stronger when the NEOS/Daya Bay data are considered,
with about 0.15\% parameter goodness of fit for all the Gallium detection cross section models,
whereas considering the NEOS/RENO data we obtain about 1.3\%.
The tensions increase significantly when we consider the combined analysis
of reactor spectral ratios and rates discussed in Section~\ref{sec:rates-ratios}.
In particular, considering the recent reactor flux models of
Estienne, Fallot \textit{et al}~\cite{Estienne:2019ujo} (EF)
and of
Kopeikin \textit{et al.}~\cite{Kopeikin:2021ugh} (KI),
the parameter goodness of fit is about 0.04\% or much lower,
depending on the consideration of the NEOS/Daya Bay or NEOS/RENO data
and a specific Gallium detection cross section model.
The minimal tension is obtained with the extreme Ground State
Gallium detection cross section model,
with the parameter goodness of fit between about 0.002\% and 0.04\%,
and the maximal tension is obtained with the Semenov model~\cite{Semenov:2020xea}
with parameter goodness of fits between about
$10^{-6}$\%
and
$5 \times 10^{-4}$\%.

In Section~\ref{sec:KATRIN},
we presented the results of the analysis of the recent
KATRIN data~\cite{KATRIN:2021uub,KATRIN:2022ith}
on the measurement of neutrino masses
through their effects on the $\beta$-decay spectrum of Tritium.
We derived limits on 3+1 active-sterile neutrino mixing
which we consider more reliable than those presented in
Ref.~\cite{KATRIN:2022ith} by the KATRIN collaboration,
because we did not consider unphysical negative mass-squared values.
We have shown that the KATRIN data exclude large values of
$\Delta m^2_{41}$,
between about a few $\text{eV}^2$ and $10^3 \, \text{eV}^2$,
for large mixing.
Moreover, in Section~\ref{sec:reactors-tritium}
we presented the results of a combined analysis of the reactor data
and the data of the Tritium experiments,
considering not only the KATRIN experiment,
but also the previous
Mainz~\cite{Kraus:2012he}
and
Troitsk~\cite{Belesev:2012hx,Belesev:2013cba},
which are relevant.
We have shown that
with this set of data,
the tension with the neutrino oscillation explanation of the Gallium Anomaly
is increased with respect to that obtained with the reactor data alone.

We presented in Section~\ref{sec:solar} the updated solar neutrino bound on the mixing of $\nu_{4}$ with $\nu_{e}$, which is in good agreement with the bound from Ref.~\cite{Goldhagen:2021kxe}.
Finally, we discuss in Section~\ref{sec:global} the results of the global fit
of the $\nu_e$ and $\bar\nu_e$ disappearance data.
We have quantified the tension of the solar neutrino bound
with the neutrino oscillation explanation of the Gallium Anomaly
and we have shown that the tension is dramatic in the case of the global fit,
with values of the parameter goodness of fit well below 1\%
for all the cases with different data choices (NEOS/Daya Bay or NEOS/RENO),
different reactor flux models, and different Gallium detection cross sections.

In conclusion, we think that the results presented in this paper
show the present status of our knowledge on
short-baseline $\nu_e$ and $\bar\nu_e$ disappearance
in the framework of 3+1 active-sterile neutrino mixing
and the dramatic tension between the neutrino oscillation explanation of the Gallium Anomaly
and the results of the other experiments.
We conclude that it is very likely that the Gallium Anomaly
is not due to neutrino oscillations and some other explanation
must be found. 

%%%%%%%%%%%%%%%%%%%%%%%%%%%%%%%%%%%%%%%%%%%%%%%%%%%%%%%%

\begin{acknowledgments}
We would like to thank:
Mikhail Danilov
and
Nataliya A. Skrobova
for sending us the DANSS $\chi^2$ table;
Kim Yeongduk, Soo-Bong Kim, Jonghee Yoo, Yoomin Oh, and all the NEOS and RENO collaborations for providing helpful information
on the NEOS+RENO data analysis;
Thierry Lasserre and Lisa Schl\"uter
for providing useful information on the KATRIN data,
and
Sanshiro Enomoto,
Leonard K\"ollenberger, and
Alexey Lokhov
for useful discussions on the KATRIN data analysis at the NuMass 2022 workshop.
C.G. and C.A.T. are supported by the research grant ``The Dark Universe: A Synergic Multimessenger Approach'' number 2017X7X85K under the program ``PRIN 2017'' funded by the Italian Ministero dell'Istruzione, Universit\`a e della Ricerca (MIUR). C.A.T. also acknowledges support from {\sl Departments of Excellence} grant awarded by MIUR and the research grant {\sl TAsP (Theoretical Astroparticle Physics)} funded by Istituto Nazionale di Fisica Nucleare (INFN).
The work of Y.F.Li and Z.Xin was supported by National Natural Science Foundation of China under Grant Nos.~12075255 and 11835013, by the Key Research Program of the Chinese Academy of Sciences under Grant No.~XDPB15.
O.T. is supported by a grant funded by Italian Ministero degli Affari Esteri e della Cooperazione Internazionale (MAECI) and also by the Indian Prime Minister's Research Fellow (PMRF) program.
\end{acknowledgments}

%%%%%%%%%%%%%%%%%%%%%%%%%%%%%%%%%%%%%%%%%%%%%%%%%%%%%%%%

\bibliographystyle{apsrev4-1}
\bibliography{main}

\end{document}